\documentclass[aps,prb,reprint,showpacs,superscriptaddress,longbibliography,floatfix]{revtex4-2}
\usepackage[table,xcdraw]{xcolor}
\usepackage{mathtools,amssymb,graphicx,units}
\usepackage{tikz}
\usetikzlibrary{shapes.geometric}
\usepackage[plainpages=false,pdfpagelabels,colorlinks=true,linkcolor=blue,urlcolor=blue,citecolor=blue,pdftitle={Critical amplitudes of the Baxter-Wu model from MCPA and Wolff simulations: free-exponent and fixed-exponent fits},pdfauthor={Vyacheslav Mozolenko, Ian Pile, Lev Shchur},pdfdisplaydoctitle=true,pdfduplex=DuplexFlipLongEdge]{hyperref}
\usepackage[english]{babel}
\usepackage{xcolor} 
\usepackage[normalem]{ulem} 

\providecommand{\RevComment}[1]{}

\newcommand{\Tc}{T_{\mathrm{c}}}
\newcommand{\abs}[1]{\left|#1\right|}
\newcommand{\avg}[1]{\langle#1\rangle}

\newcommand{\valAzeroMinus}{0.1875(26)}
\newcommand{\valAzeroPlus}{0.190(18)}
\newcommand{\valEoneMinus}{1.2810(37)}
\newcommand{\valEonePlus}{1.274(12)}
\newcommand{\valBzero}{1.2840(27)}
\newcommand{\valPzero}{1.2125(31)}
\newcommand{\valGammaMinus}{0.0677(41)}
\newcommand{\valGammaMinusP}{0.0623(20)}
\newcommand{\valGammaPlus}{0.2703(84)}
\newcommand{\valRatioA}{1.011(17)}

\newcommand{\valRatioBP}{1.05902(48)}
\newcommand{\valRatioGamma}{1.087(20)}
\newcommand{\valRchi}{3.99(12)}
\newcommand{\valRCminus}{0.00514(17)}
\newcommand{\valRCplus}{0.02073(37)}
\newcommand{\valBone}{-0.5299(91)}
\newcommand{\valPone}{-0.2343(90)}
\newcommand{\valAoneMinus}{-0.437(10)}
\newcommand{\valAonePlus}{-0.294(21)}
\newcommand{\valRCfree}{0.004967(45)}
\newcommand{\valCrossE}{0.04}
\newcommand{\valCrossMs}{0.08}
\newcommand{\valCrossPs}{0.09}
\newcommand{\valCrossC}{2.98}
\newcommand{\valCrossChi}{2.50}
\newcommand{\valMedianDevOne}{30}
\newcommand{\valMedianDevTwo}{0.55}
\newcommand{\valWorstDevOne}{73}
\newcommand{\valWorstDevTwo}{5.0}
\newcommand{\valNfits}{90}

\newcommand{\BWDataKeyRef}{Dataset symbols as in Fig.~\ref{fig:C_vs_t}.}

\begin{document}

\title{Critical amplitudes of the Baxter--Wu model from MCPA and Wolff
       simulations}

\author{Vyacheslav Mozolenko}
\affiliation{HSE University, 101000 Moscow, Russia}

\author{Ian Pil\'{e}}
\affiliation{HSE University, 101000 Moscow, Russia}

\author{Lev Shchur}
\affiliation{HSE University, 101000 Moscow, Russia}


\begin{abstract}
The two‑dimensional Baxter–Wu model belongs to the universality class of the four‑state Potts model. Exact expressions are available for its energy, specific heat, magnetization, and polarization amplitudes. We use these results as a benchmark for evaluating two simulation algorithms and two approaches to extracting critical amplitudes. Microcanonical population annealing (MCPA) and the sublattice‑freezing Wolff algorithm yield highly consistent estimates for the energy and order parameters when applied to lattices of comparable size. According to the exact Joyce series, the local effective exponent does not stabilize at a constant value but instead undergoes a monotonic drift. This behaviour stems from corrections to scaling, which can be expressed as a series in the cube root of the reduced temperature. As a result, these corrections decay relatively slowly, with the leading term being of relative order $\abs{t}^{2/3}$. When performing a fit with a free exponent on an automatically identified plateau, this drift gets incorporated into the estimated exponent, introducing a substantial bias in the derived amplitudes. In contrast, a fit that fixes the exponent to its exact value and explicitly accounts for the analytic correction series delivers a close match to the known exact amplitudes. Using this refined fitting approach, we obtain estimates for the susceptibility amplitudes and the universal amplitude ratios. These values are in good agreement with predictions from form‑factor calculations. Additionally, we numerically determine the leading correction coefficient for the polarization.
\end{abstract}

\maketitle

\section{Introduction}
\label{sec:intro}

Near a continuous phase transition, thermodynamic quantities exhibit power‑law behaviour with respect to the distance from the critical temperature $\Tc$: they either diverge or vanish accordingly. For instance, the specific heat follows the scaling form $C\simeq A_0\abs{t}^{-\alpha}$, where $t=(T-\Tc)/\Tc$ denotes the reduced temperature, $\alpha$ is the critical exponent, and $A_0$ represents the corresponding critical amplitude. The critical exponents are universal: they are identical for all systems belonging to the same universality class. In contrast, the critical amplitudes are sensitive to microscopic details and vary between different models. However, certain dimensionless combinations of these amplitudes - known as amplitude ratios - are themselves universal. Alongside the critical exponents, such ratios provide a complete characterization of the universality class~\cite{Privman1991}.

Exact amplitude ratios are known for only a limited number of models. In the case of the classic 2D square‑lattice Ising model, Wu \textit{et al.} \cite{WuMcCoyTracyBarouch1976} derived the exact spin‑spin correlation functions within the scaling region; these functions unambiguously determine the amplitudes appearing in the susceptibility expansion. For 2D  Potts model, Delfino and Cardy \cite{DelfinoCardy1998} obtained estimates for several amplitude ratios using a two‑kink form‑factor approximation. Delfino \textit{et al.} \cite{DelfinoBarkemaCardy2000} extended this calculation to the transverse susceptibility and tested the predictions with Monte Carlo simulations, and Enting and Guttmann \cite{EntingGuttmann2003} estimated the susceptibility amplitudes from series expansions. When estimating critical exponents and amplitudes with Monte-Carlo methods for systems in the four-state ($q=4$) Potts universality class, it is essential to account for logarithmic corrections to scaling. Such corrections alter the simple power‑law dependence over the temperature ranges that are practically accessible in simulations \cite{SHCHUR2010491,ShchurBerche2008, DelfinoBarkemaCardy2000,EntingGuttmann2003} and generally make research much harder. 

The Baxter–Wu (BW) model~\cite{Baxter1973,Baxter1974} shares the same universality class as the $q=4$ Potts models, but it is free of logarithmic corrections: the exact series contain only power‑law terms~\cite{Joyce1975a,Joyce1975b}, and Monte-Carlo renormalization‑group analysis confirms purely power‑law scaling~\cite{Novotny1982}. Near the critical temperature $\Tc$, the model’s energy (eq. \ref{eq:e_series}), specific heat (eq. \ref{eq:C_series}), and magnetization (eq. \ref{eq:m_series}) are described by exact series derived by Joyce~\cite{Joyce1975a,Joyce1975b}. Furthermore, the leading polarization amplitude (eq. \ref{eq:P0}) has been determined exactly~\cite{BaxterSykesWatts1975}. These exact results render the Baxter–Wu model an ideal test bed: any amplitude extracted from simulation data can be directly compared to its known exact value. For example, Shchur and Janke exploited this BW quality in Metropolis simulations and obtained numerical estimates for the susceptibility amplitudes~\cite{SHCHUR2010491}.

We employ this model to investigate two key aspects. The first is to explore the performance of simulation algorithms we used in this study, the microcanonical population annealing (MCPA) algorithm - a relatively recent method~\cite{rose2019,mozolenko2024a} - with the Wolff cluster algorithm~\cite{wolff1989} implemented in the sublattice‑freezing variant proposed by Novotny and Evertz~\cite{NovotnyEvertz1993}. These algorithms generate equilibrium data through fundamentally different mechanisms. The MCPA algorithm reconstructs the microcanonical entropy $S(E) = \ln g(E)$, where g(E) is the density of states, using a population of replicas and subsequently derives canonical averages across all temperatures via reweighting. In contrast, he Wolff algorithm explores the configuration space at a single temperature using a Markov chain. Agreement between the two methods at matched lattice sizes constitutes a non‑trivial consistency check for both algorithms. The availability of exact amplitudes further supplies an independent benchmark to validate the results of both methods.

The second question we address is the extraction of amplitudes from simulation data. Near the critical temperature $\Tc$, corrections to scaling alter the leading power‑law behaviour, which means that any amplitude estimate is sensitive to the choice of the temperature window used in the fitting procedure. In the paper \cite{SHCHUR2010491} these windows were selected manually and reported the spread of the resulting estimates as an indication of uncertainty. A more systematic, automated approach involves identifying a temperature window where the local logarithmic slope of the data remains approximately constant, and then performing a fit to extract both the amplitude and the critical exponent within that range. We refer to this approach as the free‑exponent fit. However, analysis based on the exact series for the Baxter–Wu model demonstrates that no such stable plateau exists for the local slope. Instead, the slope exhibits a monotonic drift - which happens because the exact series are expressed in powers of $\abs{t}^{1/3}$~\cite{Joyce1975a,Joyce1975b}: the leading correction term is of relative order $\abs{t}^{\theta}$ with $\theta=2/3$, and subsequent terms decrease only by factors of order $\abs{t}^{1/3}$. At $\abs{t}=10^{-2}$, this factor remains non-negligible — approximately $0.2$ - which implies that corrections decay relatively slowly and should not be omitted in typical simulation ranges.

An alternative strategy is to fix the critical exponent to its exact known value and fit the data while explicitly including the analytic correction series. This procedure is applied across a systematically scanned set of temperature windows to assess stability and accuracy. The primary objective of the present work is to quantify how trustworthy the two simulation algorithms and the two fitting procedures reproduce the exactly known amplitudes of the Baxter–Wu model. By identifying the most accurate combination of method and procedure, we aim to obtain reliable estimates for the amplitudes that are not accessible analytically and must be determined numerically. Our results indicate that the two algorithms exhibit excellent agreement for the energy and order parameters, with relative discrepancies on the order of $0.1\%$. For fluctuation observables (such as specific heat and susceptibility), the agreement remains strong, with differences at the level of a few percent. When comparing the fitted amplitudes to the exact reference values, we observe a median deviation of $17\%$ for the free‑exponent fit, whereas the fixed‑exponent fit yields a significantly smaller median deviation of $0.37\%$. Using the more stable fixed‑exponent fitting procedure, we proceed to estimate several quantities that lack exact solutions: the susceptibility amplitudes  $\Gamma_\pm$ and $\Gamma_-^{(p)}$ (eq. \ref{eq:sus_m}, \ref{eq:sus_p}), the universal ratio $R_\chi$ of the two susceptibility amplitudes, the specific‑heat combinations $R_C^\pm$ (as defined in Sec.~\ref{sec:results:ratios}), and the leading correction coefficient for the polarization.

The article is structured as follows. In Section~\ref{sec:model}, we introduce the Baxter–Wu model and lay out the principal observables under consideration. We also compile the existing exact results for the model and clarify the mechanism responsible for the drift in the effective exponent. Section~\ref{sec:methods} describes the simulation protocols and incorporates a set of validation checks to assess the quality and reliability of the data produced. Moving to Section~\ref{sec:estimators}, we formally define the two fitting procedures compared in this work and elaborate on their practical implementation, highlighting key methodological choices.  Section~\ref{sec:results} presents a head‑to‑head comparison of the simulation algorithms and fitting methods, along with a detailed discussion of the amplitude and ratio estimates they yield. Finally, Section~\ref{sec:summary} wraps up the paper by summarizing the core findings and reflecting on their broader implications for the analysis of critical amplitudes.

\section{Model, observables, and exact results}
\label{sec:model}

\subsection{Hamiltonian, sublattices, and ground states}
\label{sec:model:hamiltonian}

The Baxter--Wu model is defined on a triangular lattice of $N=L^2$ sites, where
$L$ is the linear lattice size. Each site carries an Ising spin
$\sigma_i=\pm1$. The three spins on every triangular face interact through
\begin{equation}
  \mathcal{H} = -J \sum_{\mathrm{faces}} \sigma_i\,\sigma_j\,\sigma_k ,
  \label{eq:hamiltonian}
\end{equation}
where $J>0$ is the coupling and the sum runs over all $2N$ faces. We set $J=1$
and the Boltzmann constant $k_B=1$ throughout. The energy per site
$e=\mathcal{H}/N$ then lies in $[-2,2]$.

\begin{figure}[!ht]
  \centering
  \begin{tikzpicture}[scale=0.38, every node/.style={inner sep=1.6pt, draw=black!45}]
    \foreach \i in {0,...,8} {
      \foreach \j in {0,...,8} {
        \pgfmathtruncatemacro{\sub}{mod(\i-\j+27,3)}
        \ifnum\sub=0 \def\subcol{red!75}\def\subshape{circle}\fi
        \ifnum\sub=1 \def\subcol{blue!75}\def\subshape{rectangle}\fi
        \ifnum\sub=2 \def\subcol{green!65!black}\def\subshape{regular polygon}\fi
        \node[fill=\subcol,shape=\subshape,regular polygon sides=3] at (\i + 0.5*\j, {0.8660254*\j}) {};
      }
    }
  \end{tikzpicture}
  \caption{Triangular lattice partitioned into three interpenetrating
           sublattices: red circles ($1$), blue squares ($2$), and green
           triangles ($3$). Every triangular face has one corner
           on each sublattice. Each sublattice is a triangular lattice whose
           spacing is $\sqrt{3}$ times the original one, rotated by $30^\circ$.
           The same partition defines the frozen and active sites of the cluster
           update in Sec.~\ref{sec:wolff}.}
  \label{fig:sublattices}
\end{figure}
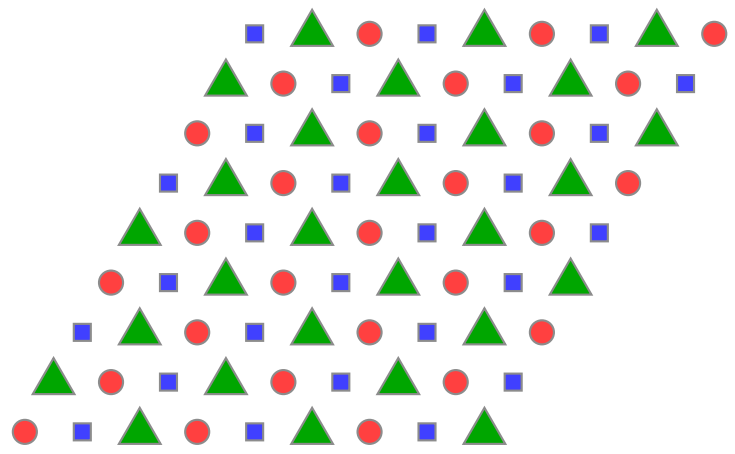

The triangular lattice splits into three interpenetrating sublattices, which we
label $1$, $2$, and $3$ (Fig.~\ref{fig:sublattices}). Each sublattice contains
$N/3$ sites, and every triangular face has one corner on each sublattice. The
energy~\eqref{eq:hamiltonian} is unchanged when the spins of any two
sublattices are reversed, because each face then contains two reversed spins.
The ground state, with $e=-2$, is therefore four-fold degenerate. In the
ferromagnetic ground state all spins are $+1$. In the three ferrimagnetic
ground states, two sublattices carry $-1$ and one carries
$+1$~\cite{Baxter1973}. The order parameters below must treat these four states
on an equal footing.

The model is self-dual~\cite{Wood1972,Merlini1972}, and the self-duality
condition fixes the exact critical temperature~\cite{Baxter1973,Baxter1974}
\begin{equation}
  \Tc = \frac{2}{\ell} = 2.269185314\ldots, \qquad \ell \equiv \ln(\sqrt{2}+1).
  \label{eq:Tc}
\end{equation}
This is also the critical temperature of the square-lattice Ising model. The
constant $\ell$ reappears in all exact amplitudes below.

\subsection{Order parameters and response functions}
\label{sec:observables}

We start from sublattice quantities. Let $\Lambda_a$ denote the set of sites of
sublattice $a\in\{1,2,3\}$. The sublattice magnetization per total site is
\begin{equation}
  m_a = \frac{1}{N}\sum_{i\in\Lambda_a}\sigma_i ,
  \label{eq:ma_def}
\end{equation}
so that $m_a=1/3$ in the ferromagnetic ground state. The sublattice polarization
measures nearest-neighbour two-spin correlations,
\begin{equation}
  p_a = \frac{1}{N}\sum_{(j,k)\in\Lambda_a}
        \sigma_{j,k}\bigl(\sigma_{j,k-1}+\sigma_{j+1,k+1}+\sigma_{j-1,k}\bigr) .
  \label{eq:pi_def}
\end{equation}
Here the lattice is stored as an $L\times L$ array with periodic boundaries, in
which the site $(j,k)$ has the six nearest neighbours $(j\pm1,k)$, $(j,k\pm1)$,
$(j+1,k+1)$, and $(j-1,k-1)$, and its sublattice is fixed by $(j+k)\bmod 3$.
The sum runs over the sites $(j,k)$ of sublattice $a$, and the three spins in
parentheses are the neighbours that lie on one of the other two sublattices. In a ground
state every face has $\sigma_i\sigma_j\sigma_k=1$, so each pair product equals
the spin on the third corner of the face. In the ground states, the
polarization therefore encodes the same information as the magnetization, but
through two-spin products. It equals $p_a=1$ in the ferromagnetic ground state.

The total magnetization and polarization are
\begin{align}
  m   &= \Bigl|\sum_{a=1}^{3} m_a\Bigr|,        &
  m_s &= \sqrt{3\sum_{a=1}^{3} m_a^{2}},        \label{eq:m_def}\\
  p   &= \frac{1}{3}\Bigl|\sum_{a=1}^{3} p_a\Bigr|, &
  p_s &= \sqrt{\frac{1}{3}\sum_{a=1}^{3} p_a^{2}} . \label{eq:p_def}
\end{align}
The factors $3$ and $1/3$ normalize the root-mean-square (RMS) quantities $m_s$
and $p_s$ to unity in every ground state. In each ground state $\abs{m_a}=1/3$,
so $\sum_a m_a^2=1/3$. Likewise $\abs{p_a}=1$, so $\sum_a p_a^2=3$. The
quantities $m$ and $p$ equal unity in the ferromagnetic ground state but $1/3$
in the three ferrimagnetic ones. During a long simulation, a finite system in
zero field wanders between all four ground-state regions. The average of $m$
then mixes them with weights set by the residence times, and its temperature
dependence is not governed by the critical amplitude alone. Novotny and
Landau~\cite{NovotnyLandau1981} introduced the RMS definitions for this reason.
We use $m_s$ and $p_s$ throughout and show $m$ only for illustration.

Let $\avg{\cdot}$ denote a canonical equilibrium average at temperature $T$. The
specific heat and the magnetic susceptibilities are~\cite{SHCHUR2010491}
\begin{align}
  C          &= \frac{N}{T^{2}}\bigl(\avg{e^{2}} - \avg{e}^{2}\bigr), \\
  T\chi_-    &= N\bigl(\avg{m_s^{2}} - \avg{m_s}^{2}\bigr), \label{eq:chi_minus}\\
  T\chi_+    &= \tfrac{1}{3}N\avg{m_s^{2}} . \label{eq:chi_plus}
\end{align}

The susceptibility $\chi_-$ is used in the ordered phase, $T<\Tc$, and $\chi_+$
in the disordered phase, $T>\Tc$. Throughout, a superscript or subscript ``$-$''
refers to the ordered phase and ``$+$'' to the disordered phase.

The polarization susceptibility in the ordered phase is
\begin{equation}
  T\chi_-^{(p)} = N\bigl(\avg{p_s^{2}} - \avg{p_s}^{2}\bigr) .
  \label{eq:chi_p_minus}
\end{equation}

The factor $1/3$ in Eq.~\eqref{eq:chi_plus} reproduces the disordered-phase
convention $T\chi_+=N\avg{m^2}$ of Ref.~\cite{SHCHUR2010491}. For two distinct
sublattices $a$ and $b$, the two-sublattice reversal of
Sec.~\ref{sec:model:hamiltonian} applied to $a$ and the remaining sublattice $c$
changes the signs of $m_a$ and $m_c$ and leaves $m_b$ unchanged. The equilibrium correlation $\avg{m_am_b}$ therefore vanishes for
$a\ne b$, and $\avg{m^2}=\sum_a\avg{m_a^2}=\avg{m_s^2}/3$. Finite samples satisfy
this identity only approximately.

\subsection{Reduced temperatures and exact amplitudes}
\label{sec:model:exact}

We use two reduced temperatures, as Ref.~\cite{SHCHUR2010491} does. The
thermal observables, the energy and the specific heat, are expanded in
$t=(T-\Tc)/\Tc$, as in Joyce's series. The magnetic observables are expanded in
$\tau=(T-\Tc)/T=1-\Tc/T$. The two variables are related by $\tau=t/(1+t)$ and
coincide to leading order near $\Tc$. When a statement applies to both kinds of
observable, we write $x$ for the distance from criticality: $x=\abs{t}$ for
thermal observables and $x=\abs{\tau}$ for magnetic ones.

Near $\Tc$, an observable $y$ behaves as
\begin{equation}
  y \simeq A\,x^{\rho}\bigl(1 + a_1x^{\theta} + \cdots\bigr) .
  \label{eq:scaling_form}
\end{equation}
The critical exponent $\rho$ is universal: it is the same for all models in the
universality class. The critical amplitude $A$ depends on the microscopic
details of the model. The correction-to-scaling exponent $\theta$ and the
correction coefficients $a_1,a_2,\dots$ describe how $y$ approaches the
asymptotic power law. For the BW model, all of these quantities are known from
the exact solution.

Baxter and Wu solved the model exactly and found that the specific heat diverges as
$C\propto\abs{t}^{-\alpha}$ with $\alpha=2/3$~\cite{Baxter1973,Baxter1974}.
Baxter, Sykes, and Watts obtained the spontaneous magnetization, which vanishes
as $\abs{\tau}^{\beta}$ with $\beta=1/12$~\cite{BaxterSykesWatts1975}. The
correlation-length exponent $\nu=2/3$ then follows from the hyperscaling
relation $2-\alpha=d\nu$ with $d=2$, and the susceptibility exponent $\gamma=7/6$
from the scaling relation $\alpha+2\beta+\gamma=2$. These values are the
exponents of the two-dimensional four-state Potts model. In renormalization-group
language they correspond to the thermal eigenvalue $y_T=1/\nu=3/2$ and the
magnetic eigenvalue $y_h=2-\beta/\nu=15/8$. The exponent formulas for the
$q$-state Potts model give the same values at
$q=4$~\cite{denNijs1979,Nienhuis1982,denNijs1983}.
The BW model therefore belongs to the four-state Potts universality class.

The leading correction-to-scaling exponent is $\theta=2/3$. It enters the exact
series below in two ways. For the specific heat, the regular background term
$A_1$ in Eq.~\eqref{eq:C_series} is a correction of relative order
$\abs{t}^{\alpha}=\abs{t}^{2/3}$ to the singular term. For the magnetization the same exponent is related to the subleading magnetic renormalization-group eigenvalue $y_3=7/8$ \cite{SHCHUR2010491}. Barber proposed this value for the BW model~\cite{Barber1976}; this value is often called the Barber ansatz now. The Monte Carlo renormalization-group study of Novotny, Landau, and Swendsen found $y_3$ consistent with it~\cite{Novotny1982}. In the operator content of the
four-state Potts class, $y_3$ belongs to the second magnetic operator~\cite{DotsenkoFateev1984,SHCHUR2010491}, whose scaling dimension is then $x_{\sigma_2}=2-y_3=9/8$. The difference of the two magnetic eigenvalues, measured in units of $y_T$, $(y_h-y_3)/y_T=(15/8-7/8)\cdot2/3=2/3$, reproduces $\theta$.

The exact amplitudes come from closed-form results. Baxter and Wu found the critical temperature and the critical energy~\cite{Baxter1973,Baxter1974}, Joyce derived the expansions of the specific heat and the magnetization near
$\Tc$~\cite{Joyce1975a,Joyce1975b}, and Baxter, Sykes, and Watts obtained the leading polarization amplitude~\cite{BaxterSykesWatts1975}.

The critical energy is $e_0=-\sqrt{2}$~\cite{Baxter1973,Baxter1974}.
Joyce's series~\cite{Joyce1975a} gives the specific heat as
\begin{equation}
  C = A_0\abs{t}^{-2/3} + A_1 + A_2\,t\abs{t}^{-2/3} + A_3\abs{t}^{2/3} + \cdots ,
  \label{eq:C_series}
\end{equation}
with the coefficients
\begin{align}
  A_0 &= \tfrac{2}{9}\,\ell^{4/3}                        = 0.187787867\ldots, \\
  A_1 &= -\tfrac{1}{2}\,\ell^{2}                         = -0.388409700\ldots, \\
  A_2 &= \tfrac{2}{27}\,\ell^{4/3}(7\sqrt{2}\,\ell - 4)  = 0.295775490\ldots, \\
  A_3 &= \tfrac{5}{9}\,\ell^{8/3}                        = 0.396723182\ldots .
  \label{eq:A_coeffs}
\end{align}
The leading amplitude $A_0$ is the same in both phases. The term $A_2$ changes
sign with $t$ and distinguishes the two phases at higher order. Relative to the
leading term, the corrections are $(A_1/A_0)\abs{t}^{2/3}$,
$\operatorname{sgn}(t)(A_2/A_0)\abs{t}$, and $(A_3/A_0)\abs{t}^{4/3}$, with
$A_1/A_0=-2.07$, $A_2/A_0=1.58$, and $A_3/A_0=2.11$. The series therefore
proceeds in steps of $\abs{t}^{1/3}$, with coefficients of order one.

Integrating $C=de/dT$ gives the energy. We define the effective energy amplitude
by dividing out the leading power,
\begin{align}
  e_{\rm eff}(t)
  &= \abs{e(t) - e_0}\,\abs{t}^{-1/3} \notag\\
  &= E_1 + E_2\abs{t}^{2/3} + E_3 t
     + E_4\abs{t}^{4/3} + \cdots .
  \label{eq:e_series}
\end{align}
Its coefficients follow exactly from those of the specific heat,
\begin{align}
  E_1 &= 3\Tc A_0, & E_2 &= \Tc A_1, \notag\\
  E_3 &= \tfrac{3}{4}\Tc A_2, & E_4 &= \tfrac{3}{5}\Tc A_3 ,
  \label{eq:E_from_A}
\end{align}
and evaluate to $E_1=\tfrac{4}{3}\ell^{1/3}=1.278376407\ldots$,
$E_2=-\ell=-0.881373587\ldots$, $E_3=0.503377048\ldots$, and
$E_4=0.540143051\ldots$, in agreement with Ref.~\cite{SHCHUR2010491}. Because $e_0$ is exact, the energy offers a test with no adjustable constant.

The RMS magnetization $m_s$ has the exact
expansion~\cite{Joyce1975b,SHCHUR2010491}
\begin{equation}
  m_s = \abs{\tau}^{1/12}\bigl(B_0 + B_1\abs{\tau}^{2/3} + B_2\abs{\tau}
        + B_3\abs{\tau}^{4/3} + \cdots\bigr)
  \label{eq:m_series}
\end{equation}
in the ordered phase, with
\begin{align}
  B_0 &= 2^{3/8}\,\ell^{1/12} = 1.283264709\ldots, \\
  B_1 &= -2^{-5/8}\,\ell^{3/4} = -0.589829210\ldots,
\end{align}
and
\begin{align}
  B_2 &= \tfrac{1}{24}\,2^{-5/8}\,\ell^{1/12}\bigl(4-\sqrt{2}\,\ell\bigr)
       = 0.073615269\ldots, \\
  B_3 &= 2^{-5/8}\,\ell^{17/12} = 0.542208470\ldots .
\end{align}
The magnetization series has the same structure as the specific-heat series,
with relative corrections in steps of $\abs{\tau}^{1/3}$. The leading amplitude $P_0$ of
the RMS polarization, defined by $p_s\simeq P_0\abs{\tau}^{1/12}$, is
known exactly~\cite{BaxterSykesWatts1975},
\begin{equation}
  P_0 = \tfrac{1}{3}\,2^{15/8}\,\ell^{1/12} = 1.209873571\ldots .
  \label{eq:P0}
\end{equation}
The ratio of the two leading amplitudes is
therefore~\cite{Joyce1975b,BaxterSykesWatts1975,SHCHUR2010491}
\begin{equation}
  \frac{B_0}{P_0} = \frac{3\sqrt{2}}{4} = 1.060660\ldots .
  \label{eq:BP_duality}
\end{equation}

The correction coefficients of the polarization are determined here numerically. We write the leading one as $P_1$, defined by
$p_s\abs{\tau}^{-1/12}=P_0+P_1\abs{\tau}^{2/3}+\cdots$. If $m_s$ and $p_s$ had the same relative corrections, their ratio would be constant and equal to $B_0/P_0$ at every temperature, which would imply $P_1=B_1P_0/B_0=-0.556$.
Whether the data follow this relation is a question for the simulations; Sec.~\ref{sec:results:corrections} answers it with the measured ratio $m_s/p_s$ and with the fitted $P_1$.

The susceptibility amplitudes are defined as
\begin{equation}
T\chi_\pm\simeq\Gamma_\pm\abs{\tau}^{-7/6}
  \label{eq:sus_m}
\end{equation}
 and
\begin{equation}
T\chi_-^{(p)}\simeq\Gamma_-^{(p)}\abs{\tau}^{-7/6}
  \label{eq:sus_p}
\end{equation}
.

They are determined numerically; Shchur and Janke estimated $\Gamma_-=0.0681(1)$, $\Gamma_-^{(p)}=0.061(1)$, and $\Gamma_+=0.265(5)$~\cite{SHCHUR2010491}. Their
polarization susceptibility is built from $p$ instead of $p_s$, so their
$\Gamma_-^{(p)}$ refers to a different observable from ours. We use
$\Gamma_-$ and $\Gamma_+$ of Ref.~\cite{SHCHUR2010491} as reference values and
show their $\Gamma_-^{(p)}$ for orientation only.

As in Ref.~\cite{SHCHUR2010491}, these amplitudes refer to $T\chi$. For
$\chi_-$ and $\chi_-^{(p)}$ this is $N$ times a variance, and for $\chi_+$ it is
$N/3$ times the second moment $\avg{m_s^2}$. The amplitude of $\chi$ itself is
smaller by the factor $\Tc=2.269$; for example, $\Gamma_-$ would change from
$0.068$ to $0.030$.

\subsection{Effective exponents and effective amplitudes}
\label{sec:noplateau}

Two derived quantities link the exact series to the fits of
Sec.~\ref{sec:estimators}. Let $y(x)$ be an observable whose leading behaviour is $y\simeq Ax^{\rho}$. For the specific heat $\rho=-\alpha$, for the
susceptibilities $\rho=-\gamma$, for $\abs{e-e_0}$ the exponent is $\rho=1-\alpha=1/3$, and for the order parameters $\rho=\beta$. The local effective exponent is the logarithmic slope
\begin{equation}
  \rho_{\rm eff}(x) = \frac{d\ln y}{d\ln x} .
  \label{eq:alpha_eff}
\end{equation}
For a pure power law, $\rho_{\rm eff}$ equals $\rho$ at every $x$. We write
$\alpha_{\rm eff}=-\rho_{\rm eff}$ for the specific heat and
$\beta_{\rm eff}=\rho_{\rm eff}$ and $\beta^{(p)}_{\rm eff}=\rho_{\rm eff}$ for $m_s$ and $p_s$. The effective amplitude divides out the exact leading power,
\begin{equation}
  y_{\rm eff} = y\,x^{-\rho} = \begin{cases}
    C\abs{t}^{2/3}            & \to A_0, \\
    \abs{e-e_0}\abs{t}^{-1/3} & \to E_1, \\
    m_s\abs{\tau}^{-1/12}     & \to B_0, \\
    p_s\abs{\tau}^{-1/12}     & \to P_0, \\
    T\chi\abs{\tau}^{7/6}     & \to \Gamma ,
  \end{cases}
  \label{eq:yeff}
\end{equation}
where the arrow gives the limit $x\to0$. We denote the five effective amplitudes by $A_{\rm eff}$, $e_{\rm eff}$, $B_{\rm eff}$, $P_{\rm eff}$, and
$\Gamma_{\rm eff}$, in the order of Eq.~\eqref{eq:yeff}; $e_{\rm eff}$ is the
quantity of Eq.~\eqref{eq:e_series}. For a pure power law, $y_{\rm eff}$ is
constant. Corrections to scaling make both $\rho_{\rm eff}$ and $y_{\rm eff}$
depend on $x$.

Differentiating Eq.~\eqref{eq:C_series} shows how fast the effective exponent of
the specific heat approaches its limit,
\begin{equation}
  \alpha_{\rm eff}(x) = \frac{2}{3}
    - \frac{2A_1}{3A_0}x^{2/3}
    - \operatorname{sgn}(t)\frac{A_2}{A_0}x
    + \mathcal{O}(x^{4/3}) .
  \label{eq:alpha_eff_series}
\end{equation}
The leading deviation from $2/3$ grows as $x^{2/3}$, with the coefficient
$-2A_1/(3A_0)=1.38$. The magnetization behaves in the same way:
Eq.~\eqref{eq:m_series} gives
$\beta_{\rm eff}=1/12+[2B_1/(3B_0)]x^{2/3}+\cdots$ with $2B_1/(3B_0)=-0.31$.
The further terms of Eq.~\eqref{eq:alpha_eff_series} are of order $x$ and
$x^{4/3}$, so successive terms differ only by a factor of order $x^{1/3}$. This
factor is $0.10$ at $x=10^{-3}$, $0.22$ at $x=10^{-2}$, and $0.37$ at
$x=5\times10^{-2}$. The corrections therefore decay slowly over the whole
accessible range; even at $x=10^{-3}$ the leading term alone shifts
$\alpha_{\rm eff}$ by $0.014$.
Table~\ref{tab:drift} shows the resulting drift in the ordered phase.

\begin{table}[!ht]
\caption{Effective exponents and effective amplitudes evaluated from the exact
         series~\eqref{eq:C_series} and~\eqref{eq:m_series} in the ordered
         phase. The limits are $\alpha=2/3$, $\beta=1/12=0.0833$,
         $A_0=0.1878$, and $B_0=1.2833$.}
\label{tab:drift}
\begin{ruledtabular}
\begin{tabular}{lcccc}
$x$                    & $10^{-4}$ & $10^{-3}$ & $10^{-2}$ & $3\times10^{-2}$ \\ \hline
$\alpha_{\rm eff}$     & $0.670$ & $0.682$ & $0.749$ & $0.866$ \\
$\beta_{\rm eff}$      & $0.0827$& $0.0804$& $0.0706$& $0.0598$ \\
$A_{\rm eff}$          & $0.1869$& $0.1836$& $0.1677$& $0.1451$ \\
$B_{\rm eff}$          & $1.2820$& $1.2775$& $1.2578$& $1.2336$ \\
\end{tabular}
\end{ruledtabular}
\end{table}

Every interval of finite width therefore shows a drift of $\alpha_{\rm eff}$
and $\beta_{\rm eff}$. Over the decade $x\in[10^{-3},10^{-2}]$, $\alpha_{\rm eff}$ changes by $0.067$, and over $[10^{-3},3\times10^{-2}]$ by $0.184$. For
$x>3\times10^{-3}$, $\beta_{\rm eff}$ differs from $1/12$ by more than $5\%$.
A plateau search in $\rho_{\rm eff}$ therefore selects windows in which the effective exponent still drifts, and the leading power law alone describes the data in none of them. The effective amplitude also drifts (Table~\ref{tab:drift}), but with the exponent fixed its
leading correction is linear in the known variable $x^{2/3}$. A straight-line fit in $x^{2/3}$ has no exponent to trade against the amplitude, and its
intercept is the critical amplitude. The fixed-exponent fit builds on this.

\section{Simulations}
\label{sec:methods}

\subsection{Lattices and datasets}
\label{sec:methods:datasets}

We use the lattice construction of Ref.~\cite{SHCHUR2010491} with periodic
boundary conditions and $L$ divisible by three. All sublattices then have $N/3$ sites, and the ground states of Sec.~\ref{sec:model:hamiltonian} fit the lattice. Table~\ref{tab:datasets} lists the simulations.

\begin{table}[!ht]
\caption{Simulations used in this work. A dataset is one algorithm at one
         lattice size. The MCPA population size is $R$, with
         $n_{\rm steps}=10$ sweeps per energy level. The Wolff runs at
         $L=162$, $243$, $363$, and $486$ use $N_{\rm meas}$ recorded
         measurements at each of 200 temperatures.}
\label{tab:datasets}
\begin{ruledtabular}
\begin{tabular}{llll}
Algorithm & $L$ & Sampling & Use \\ \hline
MCPA  & $18, 36, 72$     & $R=131072$          & finite-size check \\
MCPA  & $162$            & $R=131072$          & combined means \\
MCPA  & $243$            & $R=131072$          & comparison only \\
Wolff & $162,\dots,486$  & $N_{\rm meas}=10^5$ & combined means \\
\end{tabular}
\end{ruledtabular}
\end{table}

All comparisons use the six datasets with $L\ge162$. The combined amplitude
estimates are equal-weight means over five of them: MCPA at $L=162$ and Wolff
at all four sizes. We exclude MCPA at $L=243$ from these means because its specific-heat peak departs from the trend of the smaller sizes (Sec.~\ref{sec:methods:fss}). We use equal weights because the dataset-level uncertainties of Sec.~\ref{sec:est2} describe the sensitivity to the fit window and are not statistical errors.

\subsection{Microcanonical population annealing}
\label{sec:mcpa}

Population annealing evolves a large population of replicas, each a spin configuration, through a sequence of equilibrium ensembles~\cite{machta2010,weigel2021}. The MCPA algorithm~\cite{rose2019,mozolenko2024a,mozolenko2024b} applies this idea to the microcanonical ensemble. During cooling it lowers an energy ceiling $U$ one energy level at a time. At each step, the replicas at the current ceiling energy $U$ are removed, and the population is restored to $R$ replicas by copying randomly chosen survivors. The removed fraction $\varepsilon_U$, called
the culling factor, estimates the fraction of states at energy $U$ among all states with energy at most $U$. Each culling factor thus gives one ratio of the density of states, and the accumulated ratios give the microcanonical entropy $S(E)$ up to an additive constant~\cite{rose2019,mozolenko2024a}. After each
culling step, every replica performs $n_{\rm steps}$ sweeps of single-spin flips, each accepted if the energy stays at or below the ceiling. One sweep is $N$
attempted single-spin flips.

A cooling run covers the negative-energy half of the spectrum, which contains the ordered states. A separate heating run covers the positive-energy half in the mirror-image way: it raises an energy floor one level at a time, removes the replicas at the floor energy, and accepts only flips that keep the energy at or above the floor. The
two entropy estimates are joined in their overlap near $E=0$ after aligning their additive constants.

For each observable $O$, the replicas also give the conditional mean $O_E=\avg{O}_E$ at energy $E$. Canonical averages follow by Boltzmann reweighting,
\begin{equation}
  \avg{O}(T) = \frac{\sum_E O_E\,e^{S(E)-E/T}}{\sum_E e^{S(E)-E/T}} ,
  \label{eq:reweighting}
\end{equation}
where both sums run over the sampled energy levels. One pair of runs thus yields canonical averages at every temperature. We use $R=131072$ replicas and $n_{\rm steps}=10$ sweeps per energy level at all five sizes, $L=18$, $36$, $72$, $162$, and $243$, with one random seed per size. Each replica records the sublattice magnetizations and polarizations, from which the conditional moments of Eqs.~\eqref{eq:m_def}--\eqref{eq:p_def} are formed. We evaluate the observables
on a grid of $2\times10^4$ temperatures, $10^4$ on each side of $\Tc$, log-uniformly spaced in $\abs{t}\in[10^{-4},0.5]$.

Resampling correlates replicas that descend from the same ancestor. Such a group is called a family~\cite{weigel2021}. The usual $R^{-1/2}$
scaling of statistical errors requires effectively independent replica information~\cite{rose2019,mozolenko2024a}. Both $S(E)$ and $O_E$ carry errors, $\delta S_E$ and $\delta O_E$. To first order, the error of a canonical average at temperature $T$ is
\begin{equation}
  \delta\avg{O}_T
  = \operatorname{Cov}_T(O_E,\delta S_E)
    + \sum_E \mathcal{P}_T(E)\,\delta O_E ,
  \label{eq:mcpa_observable_error}
\end{equation}
where $\mathcal{P}_T(E)$ is the canonical probability of energy level $E$ and $\operatorname{Cov}_T$ is the covariance over this distribution. The first term
comes from the entropy and the second from the conditional means. For the energy, $O_E=E$ is exact and the second term vanishes. For $m_s$ and $p_s$,
the conditional means are estimated from family-correlated replicas, so the
second term remains. Evaluating Eq.~\eqref{eq:mcpa_observable_error} requires
the errors $\delta S_E$ and $\delta O_E$, which come from several independent
populations per size. With one population per size, we quote window-scan
uncertainties for MCPA (Sec.~\ref{sec:est2}), and
Eq.~\eqref{eq:mcpa_observable_error} serves to interpret the MCPA--Wolff
differences of Sec.~\ref{sec:results:algorithms}.

\subsection{Sublattice-freezing Wolff algorithm}
\label{sec:wolff}

To overcome the critical slowing down inherent to local-update Monte Carlo methods, we employ a cluster algorithm that updates extended groups of strongly correlated spins simultaneously~\cite{swendsen1987,wolff1989}. The standard Wolff formulation targets systems governed by pairwise spin interactions. However, the Baxter--Wu (BW) Hamiltonian involves three-spin couplings. To resolve this discrepancy, we implement the sublattice-freezing adaptation proposed by Novotny and Evertz~\cite{NovotnyEvertz1993} (see also Refs.~\cite{DengBlote2010,Fytas2026,Pile2026}). In this scheme, at every update step, one of the three sublattices is selected uniformly at random and held fixed throughout that step. The spins residing on the remaining two sublattices are designated as active and are the only ones eligible for flipping. Geometrically, these active sites constitute a honeycomb lattice.

Fixing the spins on the frozen sublattice transforms the original three-spin interaction into an effective pairwise coupling among the active spins. Specifically, consider an edge $(i,j)$ connecting two active sites. The two triangular faces adjacent to this edge each feature a single frozen corner spin, denoted $\sigma_{k_1}$ and $\sigma_{k_2}$. As shown in Ref.~\cite{Pile2026}, their collective contribution to the Hamiltonian (Eq.~\eqref{eq:hamiltonian}) reduces to a term of the form $-Js_{ij}\sigma_i\sigma_j$, where the bond parameter is defined as $s_{ij}=\sigma_{k_1}+\sigma_{k_2}$. Consequently, the active spins behave as an Ising model on a honeycomb lattice, with bond strengths constrained to the set $Js_{ij}\in\{-2J,0,2J\}$.

The cluster formation process begins with a randomly chosen active site serving as the seed. A neighbouring site $j$ of any cluster member $i$ is incorporated into the cluster with probability
\begin{equation}
  \pi_{ij} = 1 - \exp(-2J\abs{s_{ij}}/T),
\end{equation}
but only if the local alignment condition $\sigma_i\sigma_j=\operatorname{sgn}(s_{ij})$ is satisfied, i.e., the bond is energetically favourable. Edges for which $s_{ij}=0$ contribute no interaction and are never included in the cluster. Once the growth process terminates, the entire cluster is flipped as a single entity. This update procedure rigorously satisfies the condition of detailed balance and guarantees ergodicity at any finite temperature~\cite{NovotnyEvertz1993,Pile2026}. A single Wolff step is defined as one complete execution of this single-cluster update, encompassing both the selection of the frozen sublattice and the subsequent cluster flip.

Simulations were carried out for lattice sizes $L=162,243,363,486$. For each size, a distinct Markov chain was simulated at 200 temperatures: 99 values below $\Tc$, 100 above $\Tc$, and one precisely at $\Tc$ (determined to numerical precision). The temperature points are distributed logarithmically in $\abs{t}$ within the range up to $\abs{t}=0.2$. Prior to data collection, each chain underwent a thermalization period of $n_{\rm therm}=2\times10^6$ Wolff steps to ensure equilibration. Subsequently, $N_{\rm meas}=10^5$ measurements were recorded, with each measurement separated by $w=100$ Wolff steps to reduce temporal correlations. The set of recorded observables comprises the total energy, the magnetizations and polarizations for each of the three sublattices, their corresponding moments, and the derived order parameters. Pseudo-random numbers required for the stochastic updates were generated using the NumPy implementation of the Mersenne Twister MT19937 algorithm.

\subsection{Finite-size check}
\label{sec:methods:fss}

On lattices of finite extent, the sharp divergences characteristic of a continuous phase transition are smoothed out once the thermodynamic correlation length becomes comparable to the linear system size $L$. Given the scaling relation $\xi \sim \abs{t}^{-\nu}$, this rounding effect becomes significant when $\abs{t}$ falls below the threshold $\abs{t}_{\min}\sim L^{-1/\nu}=L^{-3/2}$~\cite{landau2005}. Numerically evaluating this criterion yields $\abs{t}_{\min} \approx 4.8\times10^{-4}$ for $L=162$ and $\abs{t}_{\min} \approx 9.3\times10^{-5}$ for $L=486$. These estimates represent only an order-of-magnitude guide; notably, both values lie below the lower bound $x=2\times10^{-3}$ of the fitting windows used for the fixed-exponent analysis described in Sec.~\ref{sec:est2}. Empirical observation reveals that finite-size effects manifest somewhat earlier than this simple estimate suggests: the effective energy amplitude for $L=162$ deviates from the expected scaling trend below $x\approx2$--$3\times10^{-3}$ (see Fig.~\ref{fig:energy}), which corresponds to roughly four to six times the estimated $\abs{t}_{\min}$. A similar deviation is observed for the magnetic effective amplitudes derived from MCPA data at $L=243$ (Fig.~\ref{fig:effamp}), despite the smaller intrinsic $\abs{t}_{\min}$ for this system size. A detailed discussion of the underlying causes for these departures is provided in Sec.~\ref{sec:results:limits}. It is worth noting that the adaptive plateau-search algorithm employed in the free-exponent fitting procedure is capable of selecting windows that extend to lower values of $x$ (Sec.~\ref{sec:est:example}).

\begin{figure*}[!ht]
  \centering
  \includegraphics[width=\textwidth]{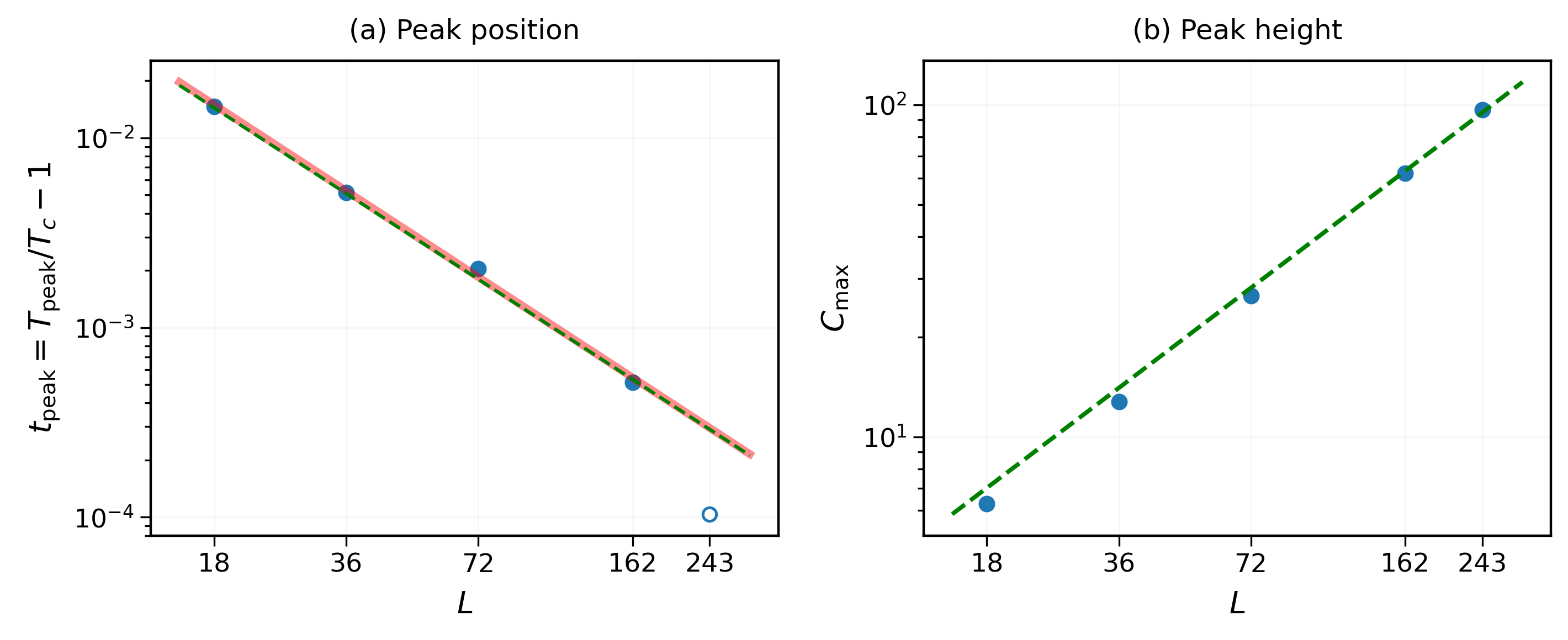}
  \caption{Scaling behaviour of the specific-heat peak: (a) positions and (b) heights, as determined by MCPA. Blue circles represent the raw MCPA data. In panel (a), the thick light-red line corresponds to an unweighted log-log fit $t_{\rm peak}=1.17L^{-1.508}$ performed over the range $L\le162$, with the open circle at $L=243$ explicitly excluded from the fit. The green dashed lines indicate the reference guides $1.1L^{-3/2}$ and $0.39L$; note that in panel (a) the guide closely aligns with the fitted curve.}
  \label{fig:fss}
\end{figure*}

To quantitatively assess the finite-size scaling behaviour, we leverage the MCPA dataset spanning lattice sizes from $L=18$ to $L=243$. According to finite-size scaling theory~\cite{landau2005}, the location of the specific-heat maximum, $T_{\rm peak}$, should converge to the critical temperature $\Tc$ such that the reduced deviation $t_{\rm peak}=(T_{\rm peak}-\Tc)/\Tc$ scales as $L^{-1/\nu}$. This follows directly from the fact that $tL^{1/\nu}$ is the appropriate scaling variable. Simultaneously, the height of the peak, $C_{\max}$, is predicted to scale as $L^{\alpha/\nu}$. For the universality class relevant to our study, this simplifies to $C_{\max}\propto L$. The numerical validation of these predictions is presented in Table~\ref{tab:peak} and Fig.~\ref{fig:fss}.

\begin{table}[!ht]
\caption{Specific-heat maximum obtained from MCPA simulations. The final column provides a direct test of the scaling hypothesis $t_{\rm peak}\propto L^{-3/2}$.}
\label{tab:peak}
\begin{ruledtabular}
\begin{tabular}{rrrrr}
$L$ & $T_{\rm peak}$ & $C_{\max}$ & $t_{\rm peak}$ & $t_{\rm peak}/L^{-3/2}$ \\
\hline
 18 & 2.302369 &  6.293 & $1.462\times10^{-2}$ & 1.117 \\
 36 & 2.280893 & 12.770 & $5.159\times10^{-3}$ & 1.114 \\
 72 & 2.273817 & 26.579 & $2.041\times10^{-3}$ & 1.247 \\
162 & 2.270353 & 62.192 & $5.146\times10^{-4}$ & 1.061 \\
243 & 2.269420 & 96.519 & $1.034\times10^{-4}$ & 0.393 \\
\end{tabular}
\end{ruledtabular}
\end{table}

Performing an unweighted log-log regression on the data for $L=18$ to $162$ yields the relation $t_{\rm peak}=1.17\,L^{-1.508}$. The fitted exponent, $-1.508$, is in excellent agreement with the theoretically expected value of $-1/\nu=-3/2$. Over this range of system sizes, the product $t_{\rm peak}L^{3/2}$ fluctuates between $1.061$ and $1.247$, implying that the leading-order scaling ansatz captures the peak positions with an accuracy of approximately $18\%$. Regarding the peak height, the ratio $C_{\max}/L$ exhibits a monotonic increase from $0.350$ at $L=18$ to $0.384$ at $L=162$, indicating that the asymptotic limit has not yet been fully reached. Consequently, we adopt $C_{\max}^{\rm FSS}=0.39L$ solely as a practical approximation for the expected peak height at those sizes.

A notable anomaly appears at $L=243$, where the scaled quantity $t_{\rm peak}L^{3/2}=0.393$ deviates from the established trend observed for smaller lattices by approximately a factor of three. Due to this significant departure, the MCPA results for $L=243$ are omitted from the calculation of combined amplitude estimates, although they are retained in the comparison tables.

\subsection{Quality of the Wolff runs}
\label{sec:wolff:quality}

Because measurements collected sequentially along a Markov chain are temporally correlated, the number of effectively independent samples in a given dataset is strictly lower than the total count $N_{\rm meas}$ of recorded observations. To rigorously quantify this reduction, we compute the integrated autocorrelation time for each measured observable. Let $\mathcal{A}_O(k)$ denote the normalized autocorrelation function of an observable $O$ evaluated at a lag of $k$ measurements; this quantity is estimated from the $N_{\rm meas}-k$ available data pairs. Following the established procedure of Sokal~\cite{Sokal1997}, the integrated autocorrelation time $\tau_{{\rm int},O}$ is estimated via the relation
\begin{equation}
  \tau_{{\rm int},O}=\tfrac12+\sum_{k=1}^{M}\mathcal{A}_O(k),
\end{equation}
where the summation cutoff $M$ is chosen as the smallest window size satisfying the condition $M\ge c_{\rm w}\tau_{{\rm int},O}(M)$. We adopt a baseline window multiplier $c_{\rm w}=6$ and further cross-validate the stability of the estimates by repeating the calculation for $c_{\rm w}=4,8,10$. The effective number of independent measurements is then obtained as $N_{\rm eff}=N_{\rm meas}/(2\tau_{\rm int})$.

As anticipated from the theory of critical slowing down, the autocorrelation times increase with system size $L$ in the vicinity of $\Tc$. Moreover, the sublattice-freezing mechanism inherently leads to slower decorrelation compared to the Fortuin--Kasteleyn bond-cluster methods applicable to pure pairwise interaction models~\cite{swendsen1987,wolff1989,Pile2026}.

\begin{table}[!ht]
\caption{Diagnostics for the Wolff simulations in the critical region $\abs{t}<0.01$, computed using the window multiplier $c_{\rm w}=6$. Here, $\tau_{\rm int}$ represents the largest autocorrelation time observed for the energy observable within this interval, expressed in units of measurements (with one measurement corresponding to $w=100$ Wolff steps). The quantity $N_{\rm eff}$ denotes the corresponding minimal effective sample size. The column $n_{\rm therm}/(w\tau_{\rm int})$ reports the thermalization duration in units of the autocorrelation time. The block length is determined separately for each temperature; the ratio block$/\tau_{\rm int}$ indicates the smallest such ratio at a given temperature, while $n_{\rm blocks}$ gives the minimal number of complete blocks. The final column compares the specific heat at $\Tc$ against the approximate MCPA peak height $C_{\max}^{\rm FSS}=0.39L$ derived in Sec.~\ref{sec:methods:fss}.}
\label{tab:wolffqc}
\begin{ruledtabular}
\begin{tabular}{rrrrrrr}
\hline
$L$ & $\tau_{\rm int}$ & $N_{\rm eff}$ & $n_{\rm therm}/(w\tau_{\rm int})$ & block$/\tau_{\rm int}$ & $n_{\rm blocks}$ & $C(T_c)/C_{\max}^{\rm FSS}$ \\
\hline
162 & 28 & 1815 & 726.2 & 4.0 & 800 & 0.76 \\
243 & 49 & 1025 & 409.9 & 4.0 & 400 & 0.74 \\
363 & 110 & 455 & 182.1 & 4.0 & 200 & 0.70 \\
486 & 204 & 245 & 98.1 & 4.0 & 100 & 0.72 \\
\hline
\end{tabular}

\end{ruledtabular}
\end{table}

Table~\ref{tab:wolffqc} provides a summary of the simulation diagnostics near the critical point. For the energy observable, the maximum autocorrelation times observed within the interval $\abs{t}<0.01$ are $28$, $49$, $110$, and $204$ measurements for lattice sizes $L=162$, $243$, $363$, and $486$, respectively. In every case, the thermalization period exceeds $98$ autocorrelation times, confirming that the system has reached a well-equilibrated state before data collection commenced.

We extend this diagnostic analysis to additional observables, including the order parameters, their statistical moments, and the centered squares that contribute to the computation of the specific heat $C$ and various susceptibilities. All examined time series comprise more than $100\tau_{\rm int}$ measurements, although some series at $L=363$ and $486$ fall short of the $1000\tau_{\rm int}$ benchmark. Sensitivity checks reveal that varying the window multiplier from $4$ to $10$ induces changes in the estimated energy autocorrelation time exceeding $25\%$ at $1$, $5$, $6$, and $24$ of the $200$ sampled temperatures for $L=162$, $243$, $363$, and $486$, respectively. Consequently, the longest autocorrelation times are associated with uncertainties of this magnitude.

To estimate the statistical error for each Wolff-based average, we employ a blocking analysis: the $N_{\rm meas}$ measurements recorded at a fixed temperature are partitioned into $n_{\rm blocks}$ consecutive blocks, and the error is inferred from the variance of the block means~\cite{Sokal1997}. The block sizes are set to be as small as $4\tau_{\rm int}$, which implies that these error estimates are inherently approximate~\cite{Sokal1997}. These errors primarily serve as a diagnostic tool to assess the reliability of the data; the actual fitting procedures, which are unweighted, utilize only the mean values (Sec.~\ref{sec:estimators}).

According to finite-size scaling theory, at the critical temperature $\Tc$ the specific heat should scale as $C(\Tc)\propto L^{\alpha/\nu}=L$, mirroring the behaviour of the peak height. Therefore, the ratio $C(\Tc)/C_{\max}^{\rm FSS}$ is expected to be approximately constant across different system sizes and strictly less than unity, since $\Tc$ occurs at a lower temperature than the position of the specific-heat peak (see Table~\ref{tab:peak}). The empirically determined values, which range between $0.70$ and $0.76$, are fully consistent with this theoretical expectation.

Far from the critical region, for instance at $\tau\simeq-0.10$, the energy autocorrelation times diminish to approximately $0.5$ measurements. At such temperatures, the four independent Wolff datasets exhibit exceptional agreement on the sublattice magnetization $m_s$, with discrepancies no larger than $2\times10^{-5}$. Collectively, these consistency checks demonstrate that the Wolff simulations are both well-equilibrated and sufficiently long to support robust amplitude estimation within the chosen fitting windows. In the critical region, the largest lattice sizes provide on the order of several hundred effectively independent measurements. It should be noted that the uncertainty associated with any fitted amplitude ultimately depends not only on $N_{\rm eff}$ but also on the specific observable under study, the details of the fitting procedure, and the particular fit being performed (Sec.~\ref{sec:results:limits}).

\section{Two approaches to amplitude estimation}
\label{sec:estimators}

Both estimation procedures operate on a common input format: sets of data pairs $(x,y)$ collected on a single side of the critical temperature $\Tc$ for a specific observable and dataset. Here, $x$ corresponds to the scaling variable defined in Sec.~\ref{sec:model:exact}, while the observable $y$ may represent the specific heat $C$, the absolute energy deviation $\abs{e-e_0}$, the sublattice magnetization $m_s$, the sublattice polarization $p_s$, or a susceptibility of the form $T\chi$. For the purpose of cross-dataset comparisons and graphical representation, we employ a standardized reference window defined as $x\in[3\times10^{-3},5\times10^{-2}]$. This interval is chosen such that its lower bound lies safely above the finite-size rounding scale $\abs{t}_{\min}$ for all considered datasets, while simultaneously spanning more than one decade in $x$, thereby capturing a broad range of scaling behaviour.

\subsection{Free-exponent fitting procedure}
\label{sec:est1}

The free-exponent fit treats the critical exponent as an unknown parameter to be determined directly from the data. It performs a linear regression on the logarithmically transformed relation
\begin{equation}
  \ln y = \ln A + \rho\ln x,
  \label{eq:est1}
\end{equation}
using unweighted least squares to estimate both the amplitude $A$ and the exponent $\rho$. The fitting window $[x_{\rm lo},x_{\rm hi}]$ is not chosen manually but is instead identified automatically as the longest contiguous interval over which the effective exponent $\rho_{\rm eff}$ (defined in Eq.~\eqref{eq:alpha_eff}) remains approximately constant. The selection algorithm proceeds through the following four stages.

First, the raw data are grouped into logarithmic bins. A resolution of $50$ bins per decade is applied to most datasets, whereas a coarser binning of $25$ bins per decade is used for the Wolff simulation data, which are already sampled on a logarithmic temperature grid. This binning step is essential to suppress the impact of numerical noise, which would otherwise dominate the local slope calculation when dealing with large datasets containing up to $10^4$ reweighted points.

Second, the effective exponent $\rho_{\rm eff}$ is computed by applying a sliding log-linear fit over a moving window of $n_w=7$ consecutive binned points; for the Wolff data, a reduced window of $5$ points is employed.

Third, every contiguous subinterval containing at least five $\rho_{\rm eff}$ values is evaluated for flatness. An interval is deemed sufficiently flat if the standard deviation of its $\rho_{\rm eff}$ values (calculated with the divisor equal to the number of points) does not exceed the tolerance threshold
\begin{equation}
  \delta = \max\bigl(2[\widetilde d + 1.4826\,\mathrm{MAD}(\{d_i\})],\;
           0.01\,\abs{\widetilde\rho_{\rm eff}}\bigr).
  \label{eq:tol}
\end{equation}
In this expression, $d_i=\abs{\rho_{{\rm eff},i+1}-\rho_{{\rm eff},i}}$ represents the absolute differences between consecutive effective-slope estimates along the entire curve, $\widetilde d$ is the median of these differences, $\mathrm{MAD}(\{d_i\})=\operatorname{median}_i\abs{d_i-\widetilde d}$ denotes their median absolute deviation, and $\widetilde\rho_{\rm eff}$ is the median value of $\rho_{\rm eff}$. The factor $1.4826$ serves to convert the median absolute deviation into an estimate of the Gaussian standard deviation, the factor $2$ sets the scale for the tolerance, and the term $0.01\abs{\widetilde\rho_{\rm eff}}$ provides a relative floor to prevent excessively tight constraints in regions of small slope. This formulation yields a robust heuristic criterion for flatness; importantly, because $\delta>0$, slowly varying, monotonic trends in $\rho_{\rm eff}$ are permitted to pass the test.

Fourth, among all intervals satisfying the flatness criterion, the one containing the largest number of data points is selected. In cases where multiple intervals contain the same number of points, the tie is resolved by choosing the interval with the smallest lower bound $x_{\rm lo}$. Once the optimal window is identified, Eq.~\eqref{eq:est1} is fitted to the original, unbinned data within this interval using unweighted least squares. If no interval meets the flatness requirement or if the best candidate contains fewer than ten data points, no amplitude estimate is reported. Crucially, this procedure imposes no prior assumptions regarding the value of the exponent $\rho$.

\subsection{Fixed-exponent fitting procedure}
\label{sec:est2}

In contrast to the free-exponent approach, the fixed-exponent fit leverages the theoretically exact critical exponent to isolate the critical amplitude. It begins by constructing the effective amplitude $y_{\rm eff}$ as defined in Eq.~\eqref{eq:yeff}. Subsequently, it models the leading corrections to scaling via a truncated series expansion:
\begin{equation}
  y_{\rm eff}(x) = \sum_{k=0}^{n_{\rm terms}-1} c_k x^{\lambda_k},
  \qquad \boldsymbol{\lambda}=(0,2/3,1,4/3),
  \label{eq:est2}
\end{equation}
where the fit is performed using unweighted least squares on $y_{\rm eff}$. The number of retained terms, $n_{\rm terms}$, is varied over the set $\{2,3,4\}$, corresponding to two-, three-, and four-term truncations. The intercept coefficient $c_0$ corresponds directly to the critical amplitude, while the higher-order coefficients $c_1,c_2,\dots$ encode the effective correction amplitudes.

The specific power sets used are $(0,2/3)$ for the two-term fit, $(0,2/3,1)$ for the three-term fit, and $(0,2/3,1,4/3)$ for the four-term fit. These exponents are rigorously grounded in the exact series expansions for energy, specific heat, and magnetization presented in Sec.~\ref{sec:model:exact}, as well as in the susceptibility analyses of Ref.~\cite{SHCHUR2010491}. For the polarization observables, the same basis set is adopted, with the correction powers inferred from the magnetization series. Each distinct susceptibility—namely $\chi_-$, $\chi_+$, and $\chi_-^{(p)}$—is fitted independently to ensure proper treatment of their individual scaling properties. Notably, in the case of the two-term fit, plotting the transformed coordinates $(x^{2/3},y_{\rm eff})$ linearizes the model, allowing the critical amplitude $c_0$ to be directly read off as the intercept at $x=0$.

To replace the manual window selection employed in Ref.~\cite{SHCHUR2010491}, we implement a systematic grid search over possible fitting windows. The lower bound $x_{\rm lo}$ is sampled at six logarithmically spaced points within the interval $[2\times10^{-3},2\times10^{-2}]$, while the upper bound $x_{\rm hi}$ takes six logarithmically spaced values in $[2\times10^{-2},2\times10^{-1}]$. This yields a total of $36$ candidate window pairs. To ensure numerical stability and physical relevance, we impose several quality criteria: each window must contain at least $12$ data points, span at least half a decade in $x$, and correspond to a design matrix with a condition number below $10^8$. After applying these filters, six of the initial $36$ windows are excluded for being too narrow, leaving $30$ viable candidates.

For each observable and dataset, these $30$ windows are combined with the three possible truncation orders ($n_{\rm terms}=2,3,4$), resulting in $90$ individual fits per observable-dataset pair. From this ensemble of estimates, we report the median value as the primary dataset-level amplitude estimate. To quantify the associated uncertainty, let $Q_{16}$ and $Q_{84}$ denote the 16th and 84th percentiles of the distribution of amplitude estimates across the $90$ fits. The quoted uncertainty is defined as the larger of the half-spread $(Q_{84}-Q_{16})/2$ and the median standard error derived from the individual fits. In practice, the half-spread consistently dominates, meaning that the reported error reflects the sensitivity of the estimate to both the choice of fitting window and the order of truncation. Because the grid search explicitly includes the narrowest windows—which tend to produce the most extreme outliers (as discussed in Sec.~\ref{sec:results:amplitudes})—this uncertainty measure is deliberately conservative.

\subsection{The two fits on the dataset}
\label{sec:est:example}

\begin{figure*}[!ht]
  \includegraphics[width=\textwidth]{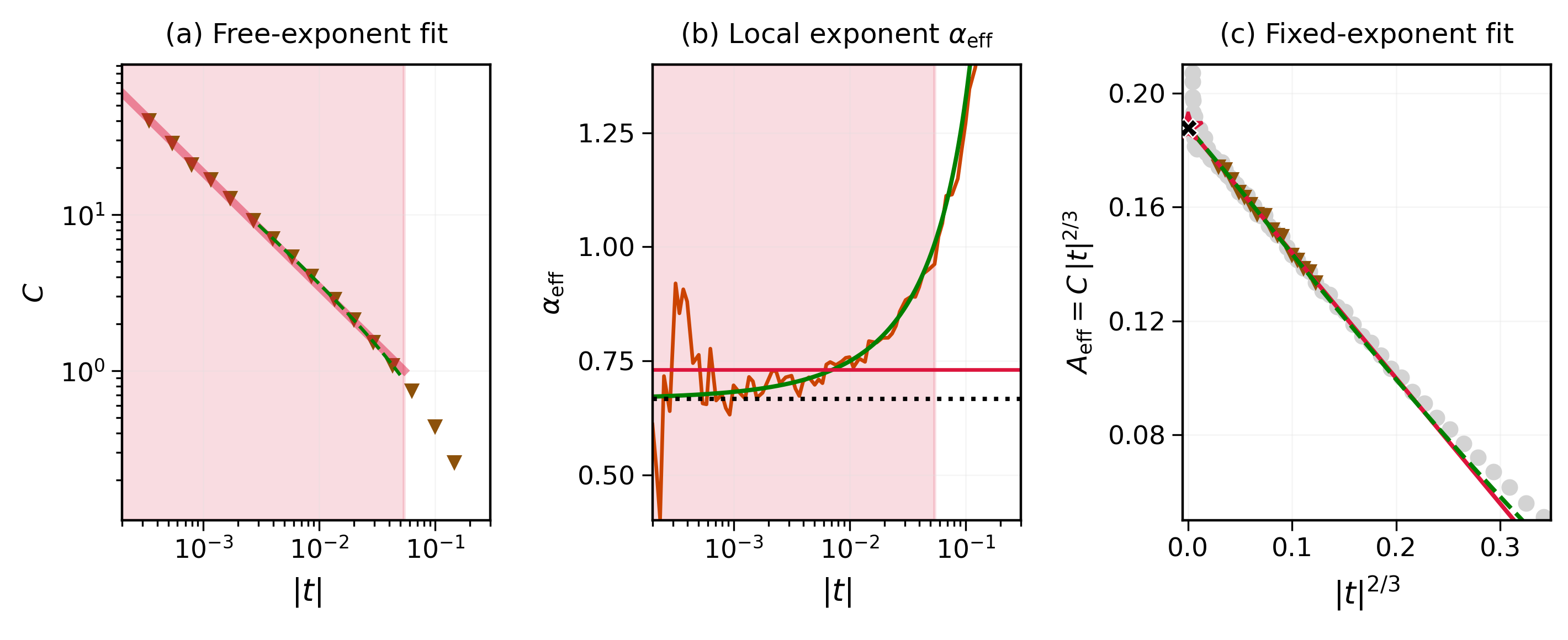}
  \caption{Comparison of the two estimation methods applied to the ordered-phase specific heat of the Wolff dataset at $L=363$. (a) Log–log representation showing the free-exponent fit (Eq.~\eqref{eq:est1}): brown downward triangles represent all data points; the thick light-crimson line depicts the free-exponent fit within the crimson-shaded, automatically selected window; the green dashed curve corresponds to the exact Joyce series. The fit yields $(\alpha,A)=(0.731,0.1193)$. (b) Effective exponent analysis: the brown curve shows the measured $\alpha_{\rm eff}$, while the green curve represents the exact theoretical prediction; the crimson shading marks the selected fitting window. The black dotted line indicates the asymptotic value $\alpha=2/3$, and the horizontal crimson line shows the fitted exponent value. (c) Representation in $(\abs{t}^{2/3},A_{\rm eff})$ coordinates suitable for the fixed-exponent method: brown downward triangles denote data points within the reference window, gray circles show points outside it; the green dashed curve is the exact Joyce series; the crimson line represents the fixed-exponent fit according to Eq.~\eqref{eq:est2}. The black X symbol marks the exact amplitude $A_0=0.18779$, overlaid on the crimson star indicating the fitted intercept.}
  \label{fig:est_explained}
\end{figure*}

Figure~\ref{fig:est_explained} demonstrates the application of both fitting methodologies to the specific heat data in the ordered phase, drawn from the Wolff simulation dataset at $L=363$. The left panel presents a log–log plot where the free-exponent fit aligns closely with the data points, yielding residuals on the order of a few percent; notably, the curvature in this representation is minimal. The central panel displays the behaviour of the effective exponent $\alpha_{\rm eff}$, highlighting the automatically identified fitting window. This window spans approximately from $2\times10^{-4}$, where statistical fluctuations render $\alpha_{\rm eff}$ noisy, up to $5\times10^{-2}$. Within this range, the exact effective exponent exhibits a gradual increase from roughly $0.67$ to approximately $1.0$. The resulting free-exponent fit yields an exponent value of $0.731$ and an amplitude of $0.1193$, which is notably $36.5\%$ lower than the theoretically exact amplitude $A_0=0.1878$.

The observed systematic bias stems from a geometric effect inherent to the fitting procedure. Within the selected window, the slowly increasing $\alpha_{\rm eff}$ is effectively replaced by its average value, which exceeds the exact asymptotic exponent $2/3$. In the context of Eq.~\eqref{eq:est1}, the parameters $\ln A$ and $\alpha=-\rho$ exhibit strong anticorrelation. Specifically, if the fitted exponent increases by an amount $\Delta\alpha$ while the fitted line continues to pass through the data cloud, the intercept $\ln A$ decreases by $\Delta\alpha\abs{\overline{\ln x}}$, where the overline denotes the average over the data points included in the fit. Given that $\abs{\overline{\ln x}}$ amounts to several units in this case, even a modest error in the exponent—such as $\Delta\alpha \approx 0.06$—can induce a substantial reduction in the estimated amplitude, on the order of tens of percent.

The right panel of Fig.~\ref{fig:est_explained} illustrates the same dataset in the coordinate system appropriate for the fixed-exponent fit. In these $(\abs{t}^{2/3},A_{\rm eff})$ coordinates, the data points fall nearly onto a straight line, and the intercept of the fitted line agrees well with the exact amplitude $A_0$. This comparison underscores how fixing the theoretically correct exponent mitigates the geometric bias that affects the free-exponent approach.

\section{Results}
\label{sec:results}

\subsection{Comparison of simulation data with the exact series expansions}
\label{sec:results:raw}

\begin{figure*}[!ht]
  \includegraphics[width=\textwidth]{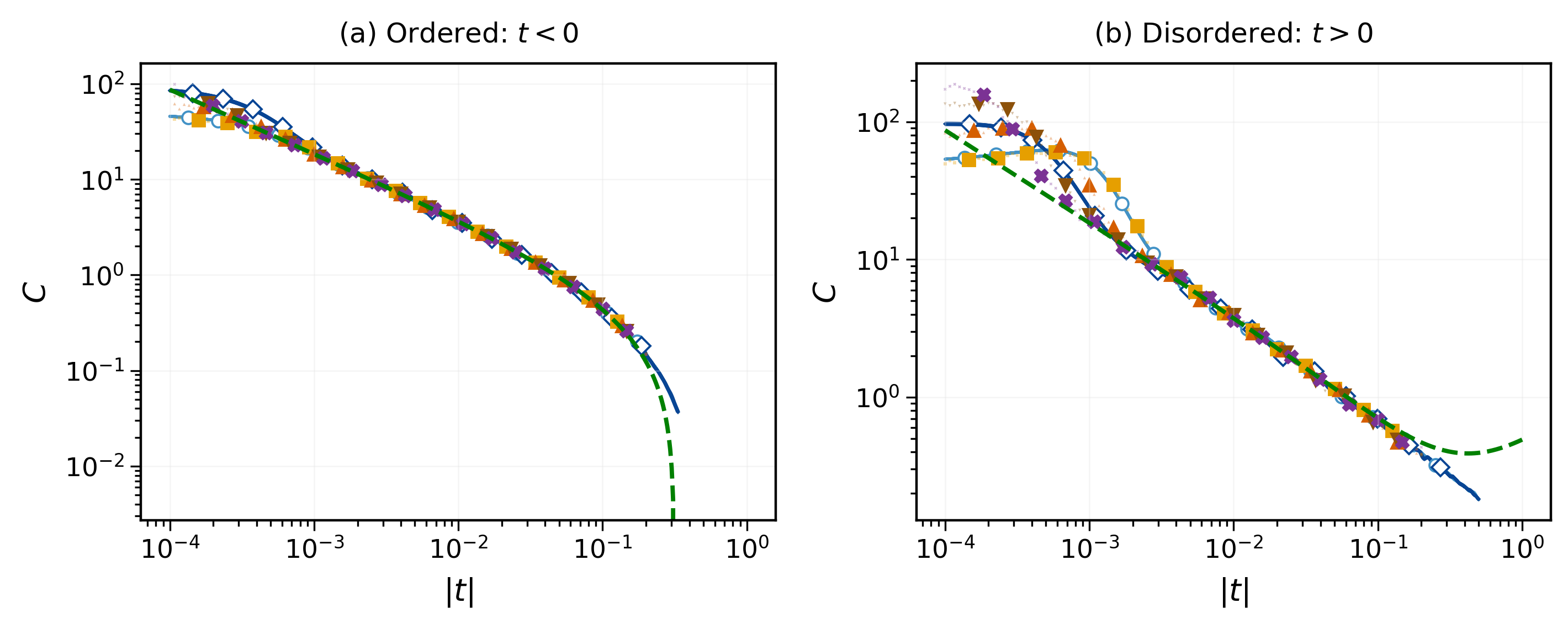}
  \caption{Specific heat $C$ versus $\abs{t}$ in both phases for all six datasets. MCPA results are shown as open blue circles ($L=162$) and open dark-blue diamonds ($L=243$). Wolff simulation data are represented by orange squares ($L=162$), vermilion upward triangles ($L=243$), brown downward triangles ($L=363$), and purple crosses ($L=486$). The green curves correspond to the four-term Joyce series given by Eq.~\eqref{eq:C_series}. For clarity, larger symbols indicate a subsampled set of points, while smaller, fainter markers represent the remaining original data points.}
  \label{fig:C_vs_t}
\end{figure*}

\begin{figure*}[!ht]
  \includegraphics[width=0.975\textwidth]{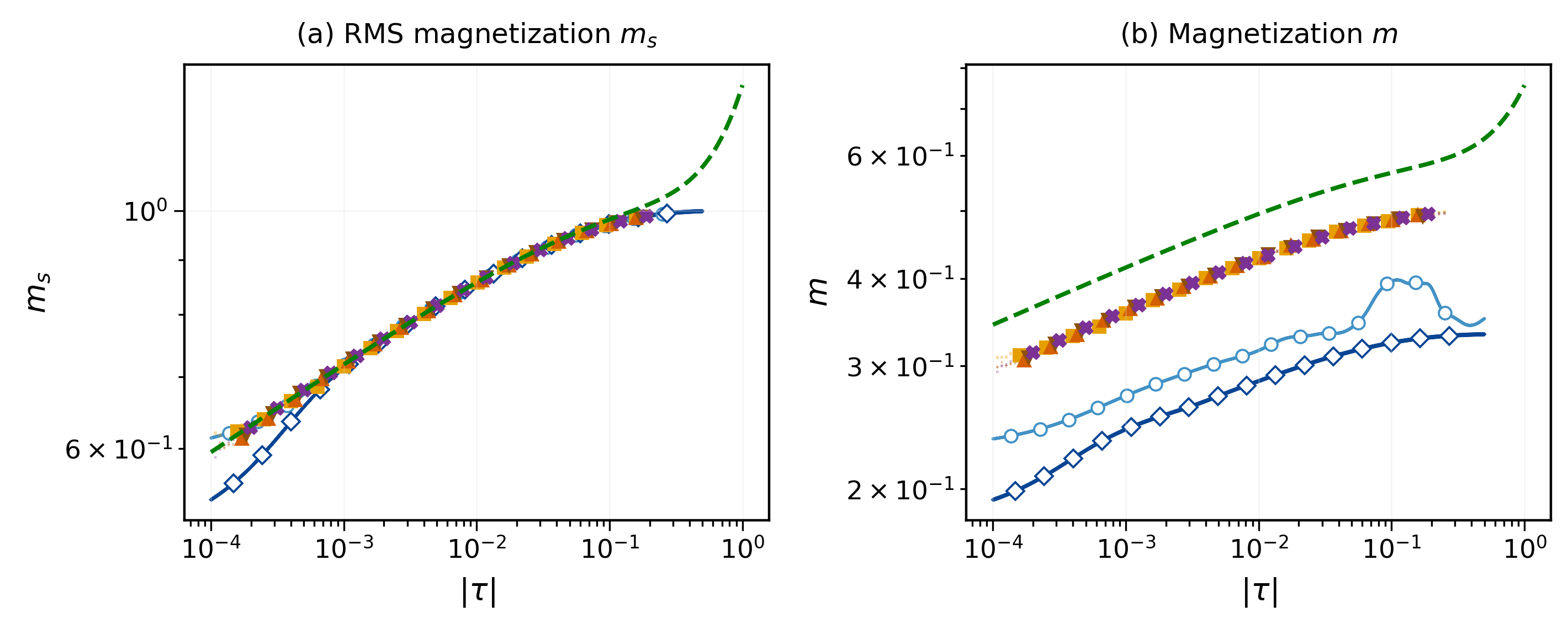}
  \caption{Magnetization in the ordered phase. (a) Root-mean-square sublattice magnetization $m_s$ plotted alongside the green four-term Joyce series (Eq.~\eqref{eq:m_series}). (b) Total magnetization $m$, which incorporates contributions from both the ferromagnetic ($m=1$) and ferrimagnetic ($m=1/3$) ground-state regions. The green dashed curve shows the same series scaled by a factor of $1/\sqrt{3}$, consistent with the relation $\avg{m^2}=\avg{m_s^2}/3$ as detailed in Sec.~\ref{sec:observables}. \BWDataKeyRef}
  \label{fig:m_vs_tau}
\end{figure*}

\begin{figure}[!ht]
  \centering
  \includegraphics[width=\columnwidth]{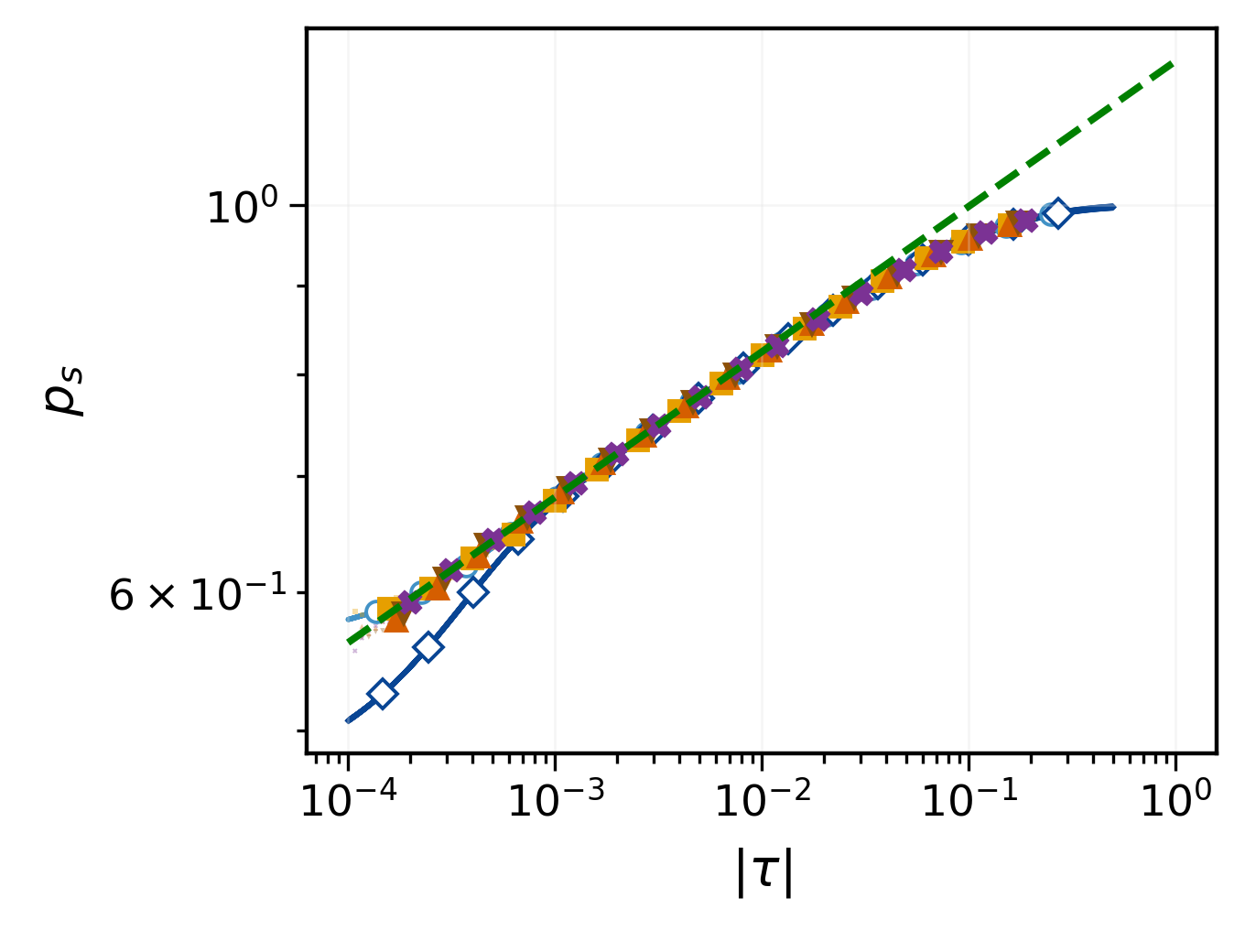}
  \caption{Root-mean-square sublattice polarization $p_s$ compared to the leading-order theoretical prediction $P_0\abs{\tau}^{1/12}$, represented by the green curve. \BWDataKeyRef\ The simulation data lie below this leading term by approximately $1$–$3\%$ in the interval $10^{-2}\lesssim\abs{\tau}\lesssim5\times10^{-2}$, with the discrepancy growing at larger values of $\abs{\tau}$. This systematic deficit is indicative of a negative correction coefficient $P_1$, as further discussed in Sec.~\ref{sec:results:corrections}.}
  \label{fig:p_vs_tau}
\end{figure}

We commence our analysis by directly comparing the raw simulation observables against the exact series expansions derived in Sec.~\ref{sec:model:exact}. Figure~\ref{fig:C_vs_t} presents the specific heat as a function of $\abs{t}$ for both phases across all six datasets. Figure~\ref{fig:m_vs_tau} illustrates the magnetization behaviour, while Fig.~\ref{fig:p_vs_tau} focuses on the polarization.

\begin{figure*}[!ht]
  \includegraphics[width=\textwidth]{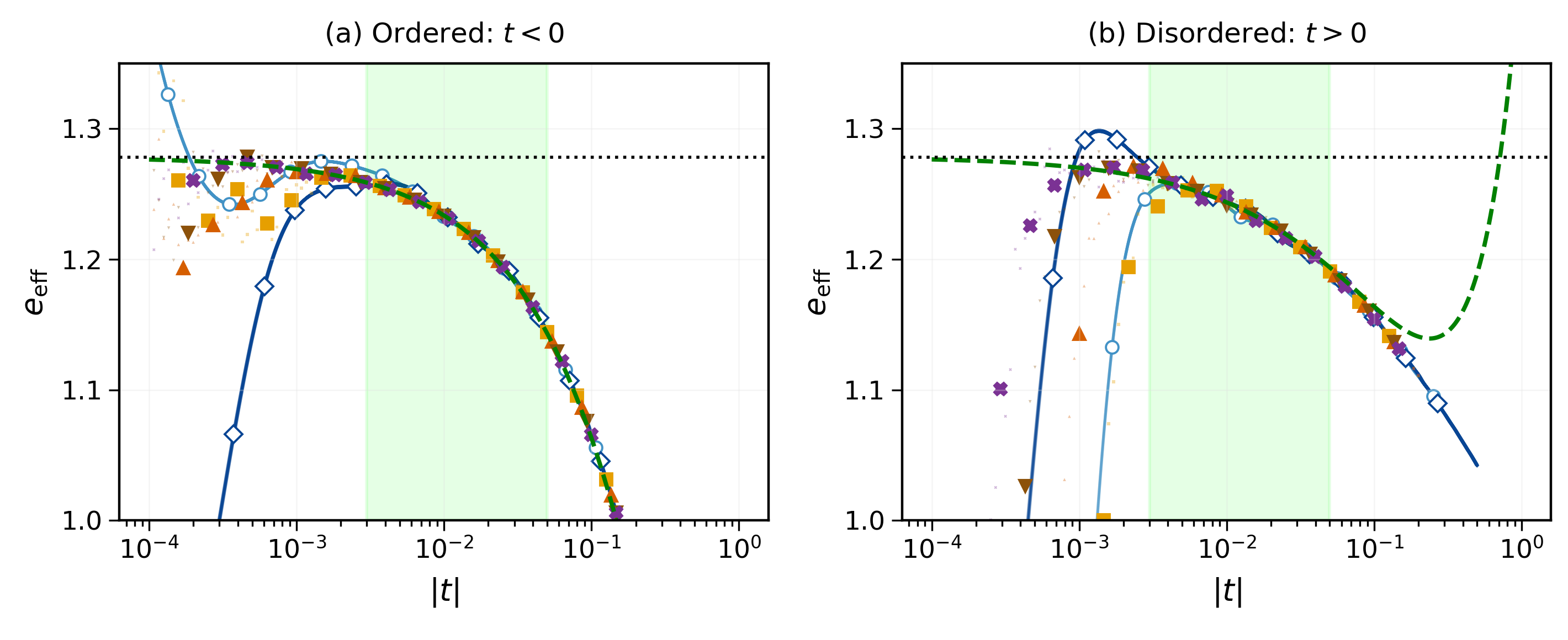}
  \caption{Effective energy amplitude defined as $e_{\rm eff}=\abs{e-e_0}\abs{t}^{-1/3}$ plotted for both phases, using the exact critical energy $e_0=-\sqrt{2}$. \BWDataKeyRef\ The green dashed curves represent the four-term series expansion (Eq.~\eqref{eq:e_series}); the black dotted line marks the exact value $E_1=3\Tc A_0=1.27838$. The green shaded band highlights the reference fitting window $0.003\le\abs{t}\le0.05$.}
  \label{fig:energy}
\end{figure*}

Among all observables, the internal energy provides the most stringent test of the theoretical series, owing to two key features: the critical energy $e_0=-\sqrt{2}$ is known exactly, and the expansion in Eq.~\eqref{eq:e_series} contains no free parameters. Figure~\ref{fig:energy} confirms that the measured effective amplitude $e_{\rm eff}$ converges to the exact coefficient $E_1=1.27838$ from both phases across all six datasets, down to $\abs{t}\approx2\times10^{-3}$. Below this threshold, finite-size effects become dominant and cause deviations from the infinite-system behaviour. Table~\ref{tab:amplitudes} summarizes the intercept values obtained from the fits.

\begin{figure*}[!ht]
  \includegraphics[width=\textwidth]{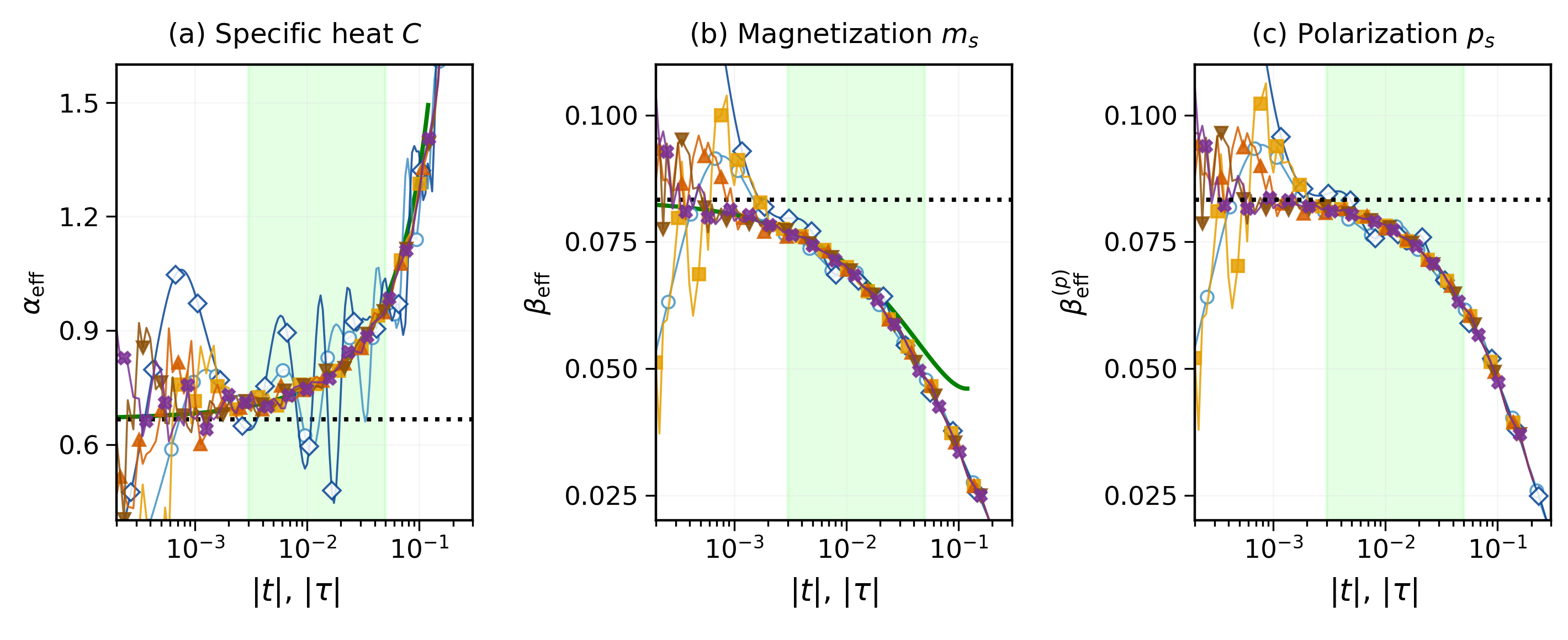}
  \caption{Local effective exponents in the ordered phase for (a) the specific heat, (b) the sublattice magnetization $m_s$, and (c) the sublattice polarization $p_s$. \BWDataKeyRef\ Green curves depict the ordered-phase Joyce series predictions for $C$ and $m_s$. Black dotted lines indicate the asymptotic critical exponents $2/3$ and $1/12$, respectively. The green shaded band marks the reference fitting window $0.003\le x\le0.05$. The observed curves systematically drift across this window, demonstrating the absence of a stable plateau.}
  \label{fig:noplateau}
\end{figure*}

Figure~\ref{fig:noplateau} offers a detailed comparison between the measured local effective exponents and their exact theoretical counterparts. For the specific heat, the Wolff simulation data closely follow the exact $\alpha_{\rm eff}$ within the reference window. In contrast, the MCPA-derived curves exhibit oscillatory behaviour around the exact prediction, with deviations reaching up to approximately $0.3$. For the magnetization, all datasets agree with the exact $\beta_{\rm eff}$ to within about $0.003$ over the range $\abs{\tau}\approx3\times10^{-3}$ to $2\times10^{-2}$. At values of $\abs{\tau}$ above $2\times10^{-2}$, the measured $\beta_{\rm eff}$ falls below the prediction of the four-term series. Conversely, below $10^{-3}$, the process of logarithmic differentiation amplifies statistical noise, leading to increased scatter in the curves. Similarly, the effective exponent for the polarization, $\beta^{(p)}_{\rm eff}$, displays a systematic drift across the reference window. Consistent with the expectations outlined in Sec.~\ref{sec:noplateau}, none of the experimentally derived curves exhibit a flat plateau, reflecting the persistent influence of correction-to-scaling terms.

\subsection{Comparison of MCPA and Wolff results at matched system sizes}
\label{sec:results:algorithms}

To facilitate a direct comparison between the two simulation approaches, both the MCPA and Wolff algorithms were executed at identical lattice sizes, specifically $L=162$ and $L=243$. Within the reference window of the ordered phase, each observable from the Wolff simulations was linearly interpolated onto the temperature grid defined by the MCPA data. Subsequently, the root-mean-square relative difference between the two sets of results was computed. The resulting discrepancies are $\valCrossE\%$ for the energy, $\valCrossMs\%$ for the sublattice magnetization $m_s$, $\valCrossPs\%$ for the sublattice polarization $p_s$, $\valCrossC\%$ for the specific heat $C$, and $\valCrossChi\%$ for the susceptibility $T\chi_-$. These values correspond to the larger difference observed across the two lattice sizes.

The analysis reveals that the energy and the primary order parameters ($m_s$ and $p_s$) exhibit excellent agreement, with relative differences below $10^{-3}$. In contrast, the fluctuation-based observables (specific heat and susceptibility) show larger discrepancies, on the order of a few percent. This pattern can be partially explained by the error structure inherent to the MCPA method, as described in Eq.~\eqref{eq:mcpa_observable_error}. For the energy, only the entropy-related error component contributes to the uncertainty. For $m_s$ and $p_s$, an additional contribution arises from the family-correlated conditional means, which is consistent with their observed differences being roughly twice as large as those for the energy. The situation is more pronounced for the specific heat and $\chi_-$, as these quantities are computed as the difference between two nearly equal moments ($\avg{y^2}-\avg{y}^2$) in the vicinity of $\Tc$. Consequently, even a small relative error in either of the individual moments can be significantly amplified in the final difference, leading to larger relative deviations. This interpretation strongly suggests that the observed differences are predominantly attributable to MCPA-specific errors rather than intrinsic inaccuracies in the Wolff algorithm.

Supporting this conclusion, the Wolff-derived amplitudes for $E_1^{(-)}$, $B_0$, and $P_0$ deviate from their exact theoretical values by at most $0.18\%$, as detailed in Sec.~\ref{sec:results:amplitudes}. A definitive separation of the error contributions from each algorithm, as well as a disentanglement of the various MCPA error sources, would require additional independent simulation runs. The oscillatory behaviour of the MCPA $\alpha_{\rm eff}$ curves shown in Fig.~\ref{fig:noplateau} further corroborates this interpretation, demonstrating similar systematic effects at the level of local effective exponents.

\subsection{Comparative analysis of the two fitting methodologies}
\label{sec:results:amplitudes}

\begin{figure*}[!ht]
  \includegraphics[width=0.98\textwidth]{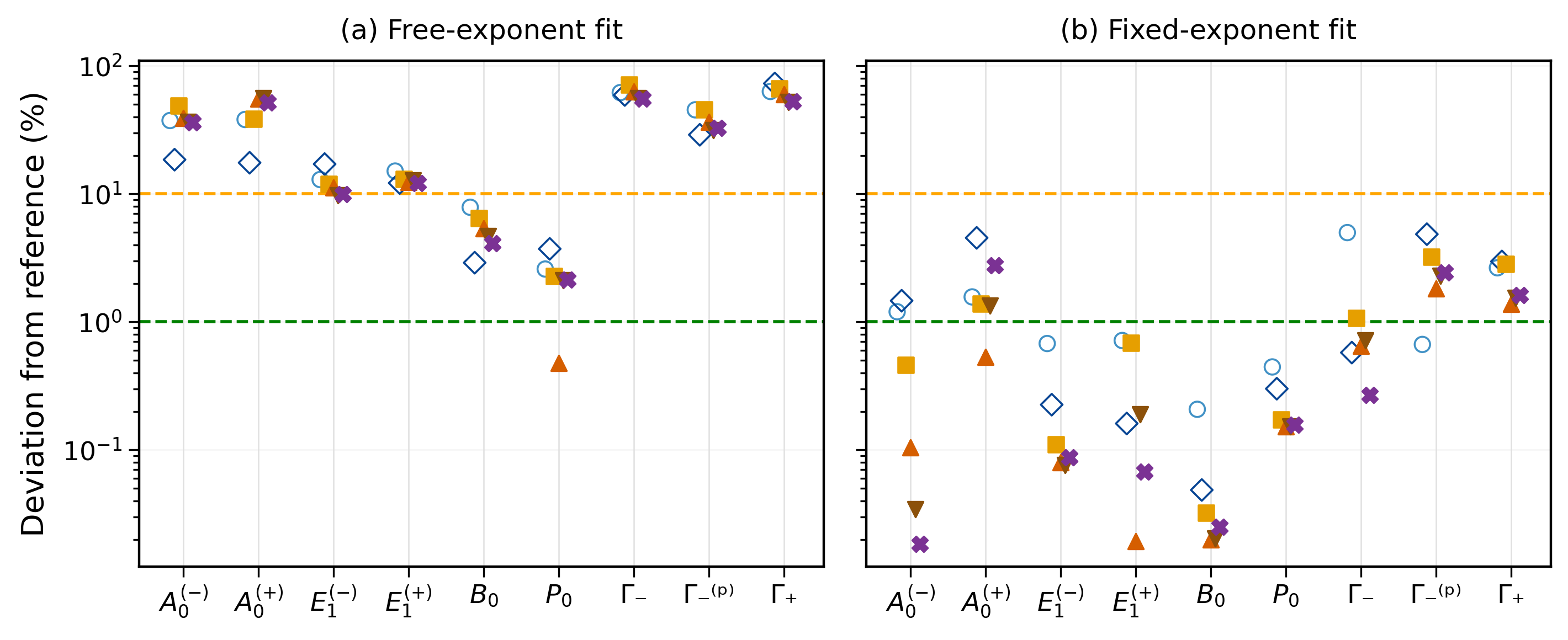}
  \caption{Relative deviations (\%) of the fitted amplitudes from their respective reference values, comparing both fitting methods applied to identical datasets. Panel (a) shows results for the free-exponent fit, while panel (b) displays those for the fixed-exponent fit. \BWDataKeyRef\ Green dashed and orange dashed horizontal lines indicate deviation thresholds of $1\%$ and $10\%$, respectively. The median and worst-case deviations are $\valMedianDevOne\%$ and $\valWorstDevOne\%$ for the free-exponent fit, and $\valMedianDevTwo\%$ and $\valWorstDevTwo\%$ for the fixed-exponent fit. These statistics include $\Gamma_-^{(p)}$, whose reference value pertains to the polarization $p$. The reference values are exact for $A_0$, $E_1$, $B_0$, and $P_0$, and are taken from Ref.~\cite{SHCHUR2010491} for the susceptibility amplitudes. The plotted deviations represent systematic offsets from these references and should not be interpreted as statistical errors.}
  \label{fig:comparison}
\end{figure*}

A comprehensive comparison of the amplitude estimates obtained via the two fitting approaches is presented in Tables~\ref{tab:amplitudes} and~\ref{tab:comparison}, while Fig.~\ref{fig:comparison} visualizes the relative deviations from the reference values. Table~\ref{tab:amplitudes} lists the amplitudes derived from the fixed-exponent fits, including their associated dataset-level uncertainties, which correspond to the half-spreads obtained from the window-scan procedure described in Sec.~\ref{sec:est2}. Table~\ref{tab:comparison} provides the corresponding free-exponent amplitude estimates for the same datasets; notably, the free-exponent fit does not employ a window-scan, and thus no such uncertainty metric is reported for these values. The reference row in both tables repeats the exact theoretical values for $A_0$, $E_1$, $B_0$, and $P_0$, along with the susceptibility amplitudes $\Gamma_\pm$ and $\Gamma_-^{(p)}$. It is important to note that the S\&J value for $\Gamma_-^{(p)}$ corresponds to the polarization $p$, not the sublattice polarization $p_s$.

\begin{table*}[!ht]
\caption{Amplitudes obtained from fixed-exponent fits. The reported uncertainties represent the dataset-level window-scan half-spreads as defined in Sec.~\ref{sec:est2}. The reference row provides the exact values for $A_0$, $E_1$, $B_0$, and $P_0$, as well as the susceptibility amplitudes~\cite{SHCHUR2010491}. Note that the $\Gamma_-^{(p)}$ value corresponds to the observable $p$ rather than $p_s$.}
\label{tab:amplitudes}
\begin{ruledtabular}
\setlength{\tabcolsep}{3pt}\footnotesize
\begin{tabular}{lrrrrrrrrr}
\hline
Dataset & $A_0^{(-)}$ & $A_0^{(+)}$ & $E_1^{(-)}$ & $E_1^{(+)}$ & $B_0$ & $P_0$ & $\Gamma_-$ & $\Gamma_-^{(p)}$ & $\Gamma_+$ \\
\hline
MCPA $L=162$ & 0.1855(90) & 0.191(44) & 1.287(12) & 1.269(19) & 1.2859(52) & 1.2153(23) & 0.0647(51) & 0.0614(47) & 0.272(20) \\
MCPA $L=243$ & 0.191(18) & 0.179(12) & 1.2813(96) & 1.280(19) & 1.2826(42) & 1.2135(46) & 0.0677(70) & 0.0640(58) & 0.257(20) \\
Wolff $L=162$ & 0.1887(28) & 0.190(30) & 1.2798(19) & 1.270(19) & 1.2837(25) & 1.2120(26) & 0.0688(39) & 0.0630(14) & 0.272(19) \\
Wolff $L=243$ & 0.1880(18) & 0.1888(97) & 1.2794(14) & 1.279(12) & 1.2835(27) & 1.2117(26) & 0.0685(35) & 0.0621(19) & 0.2687(72) \\
Wolff $L=363$ & 0.1877(24) & 0.185(14) & 1.2794(16) & 1.276(10) & 1.2835(24) & 1.2117(26) & 0.0686(37) & 0.0624(17) & 0.2691(82) \\
Wolff $L=486$ & 0.1878(22) & 0.193(18) & 1.2795(13) & 1.2775(93) & 1.2836(24) & 1.2118(26) & 0.0679(34) & 0.0625(21) & 0.2693(48) \\
\hline
reference & 0.1878 & 0.1878 & 1.2784 & 1.2784 & 1.2833 & 1.2099 & 0.0681 & 0.0610 & 0.2650 \\
\hline
\end{tabular}
\end{ruledtabular}
\end{table*}

\begin{table*}[!ht]
\caption{Free-exponent fit amplitudes computed on the same datasets used for Table~\ref{tab:amplitudes}. No window-scan uncertainty is provided for the free-exponent method. The reference row repeats the values given in Table~\ref{tab:amplitudes}.}
\label{tab:comparison}
\begin{ruledtabular}
\setlength{\tabcolsep}{3pt}\footnotesize
\begin{tabular}{lrrrrrrrrr}
\hline
Dataset & $A_0^{(-)}$ & $A_0^{(+)}$ & $E_1^{(-)}$ & $E_1^{(+)}$ & $B_0$ & $P_0$ & $\Gamma_-$ & $\Gamma_-^{(p)}$ & $\Gamma_+$ \\
\hline
MCPA $L=162$ & 0.1175 & 0.1162 & 1.1129 & 1.0855 & 1.1824 & 1.1785 & 0.0260 & 0.0333 & 0.4323 \\
MCPA $L=243$ & 0.1530 & 0.1549 & 1.0597 & 1.1228 & 1.2460 & 1.1648 & 0.0277 & 0.0433 & 0.4582 \\
Wolff $L=162$ & 0.0959 & 0.1158 & 1.1258 & 1.1115 & 1.2011 & 1.1825 & 0.0196 & 0.0332 & 0.4403 \\
Wolff $L=243$ & 0.1143 & 0.0836 & 1.1356 & 1.1199 & 1.2140 & 1.2157 & 0.0250 & 0.0388 & 0.4233 \\
Wolff $L=363$ & 0.1193 & 0.0832 & 1.1543 & 1.1158 & 1.2235 & 1.1842 & 0.0300 & 0.0418 & 0.4034 \\
Wolff $L=486$ & 0.1199 & 0.0906 & 1.1517 & 1.1230 & 1.2307 & 1.1841 & 0.0302 & 0.0411 & 0.4047 \\
\hline
reference & 0.1878 & 0.1878 & 1.2784 & 1.2784 & 1.2833 & 1.2099 & 0.0681 & 0.0610 & 0.2650 \\
\hline
\end{tabular}
\end{ruledtabular}
\end{table*}

When evaluating the overall performance of the two methods, it is instructive to consider the statistical distribution of deviations. If the entry for $\Gamma_-^{(p)}$ is excluded (since its reference value corresponds to a different physical observable, as clarified in Sec.~\ref{sec:model:exact}), the analysis of the remaining 48 entries in Tables~\ref{tab:amplitudes} and~\ref{tab:comparison} reveals a median deviation of $17\%$ for the free-exponent fit and only $0.37\%$ for the fixed-exponent fit. Including $\Gamma_-^{(p)}$ in the calculation (as done in Fig.~\ref{fig:comparison}) yields the statistics $\valMedianDevOne\%$ and $\valMedianDevTwo\%$, respectively.

The free-exponent fit exhibits a systematic bias, and the magnitude of this bias is closely correlated with the size of the leading correction-to-scaling term. For the energy, specific heat, and magnetization amplitudes, the exact relative correction coefficients are known to be $A_1/A_0=-2.07$, $E_2/E_1=-0.69$, and $B_1/B_0=-0.46$. For the polarization, a fitted value of $P_1/P_0\approx-0.19$ is obtained (see Sec.~\ref{sec:results:corrections}). The corresponding deviations observed in the free-exponent fits are substantial: $18$–$56\%$ for $A_0$, $10$–$17\%$ for $E_1$, $3$–$8\%$ for $B_0$, and $0.5$–$4\%$ for $P_0$. The susceptibility amplitudes $\Gamma_\pm$ show even larger deviations, ranging from $52$ to $73\%$; for these quantities, the correction coefficients are determined empirically from the fits themselves. This systematic trend indicates that the free-exponent fit performs worst precisely in those cases where the effective exponent $\rho_{\rm eff}$ drifts most significantly across the fitting window, despite the fact that the log–log residuals for these fits typically remain at the percent level.

In contrast, the fixed-exponent fit demonstrates superior accuracy and stability. The worst deviation observed for this method is $\valWorstDevTwo\%$. The largest deviations in this case occur for $A_0^{(+)}$ and the susceptibility amplitudes, which are also identified as the least stable estimates (as discussed in Sec.~\ref{sec:results:limits}). Notably, for the ordered-phase energy amplitude $E_1^{(-)}$ and the order-parameter amplitudes $B_0$ and $P_0$ derived from the Wolff datasets, the deviations are remarkably small, not exceeding $0.18\%$.

\begin{figure*}[tp]
  \includegraphics[width=\textwidth]{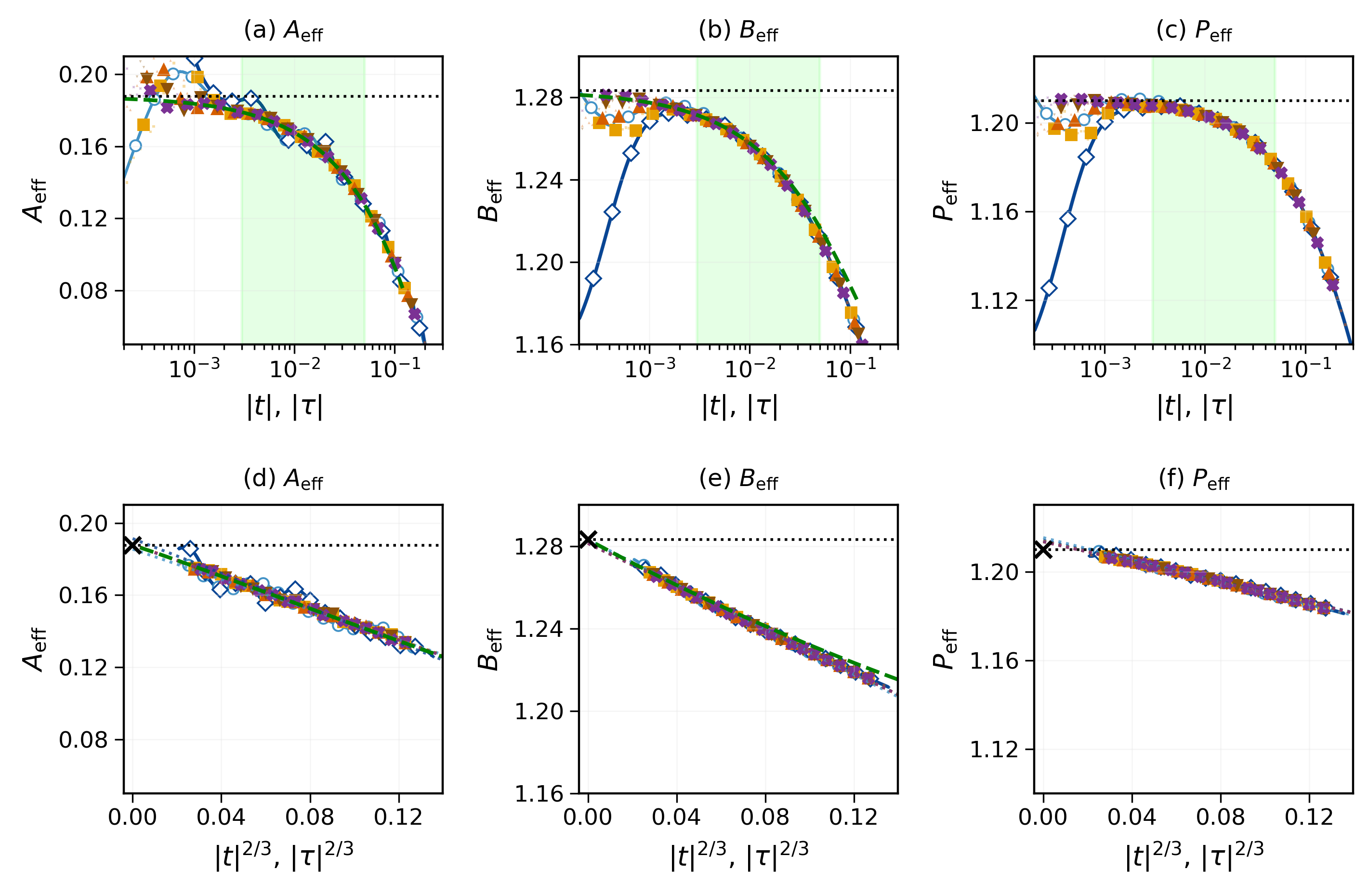}
  \caption{Effective amplitudes: (a) $A_{\rm eff}=C|t|^{2/3}$, (b) $B_{\rm eff}=m_s|\tau|^{-1/12}$, and (c) $P_{\rm eff}=p_s|\tau|^{-1/12}$. Panels (a)--(c) plot these quantities against $|t|$ or $|\tau|$ on a logarithmic axis, while panels (d)--(f) show them against $|t|^{2/3}$ or $|\tau|^{2/3}$. \BWDataKeyRef\ In panel (a), the MCPA curve for $L=243$ extends beyond the plotted axes at small $|t|$. The green shaded band in the top row marks the reference fitting window $0.003\le |t|,|\tau|\le0.05$. Green dashed curves represent the exact Joyce series for $A_{\rm eff}$ and $B_{\rm eff}$; for $P_{\rm eff}$, the exact leading amplitude $P_0$ is shown. Black dotted lines indicate exact asymptotic limits in all panels, and black crosses mark the exact intercept values in the bottom row. Dotted linear fits in the bottom row are computed using the corresponding dataset-specific data points.}
  \label{fig:effamp}
\end{figure*}

Figure~\ref{fig:effamp} illustrates the underlying behaviour of the effective amplitudes that leads to these systematic biases. It is evident that these amplitudes vary appreciably across the reference fitting window. According to Table~\ref{tab:drift}, the effective amplitude $A_{\rm eff}$ decreases by $21\%$, and $B_{\rm eff}$ by $3.4\%$, between $x=10^{-3}$ and $x=3\times10^{-2}$. In the coordinate system appropriate for the fixed-exponent fit (bottom row of Fig.~\ref{fig:effamp}), the data from all six datasets follow a common linear trend, and the fitted intercepts align closely with the exact theoretical values.

\begin{figure*}[tp]
  \includegraphics[width=\textwidth]{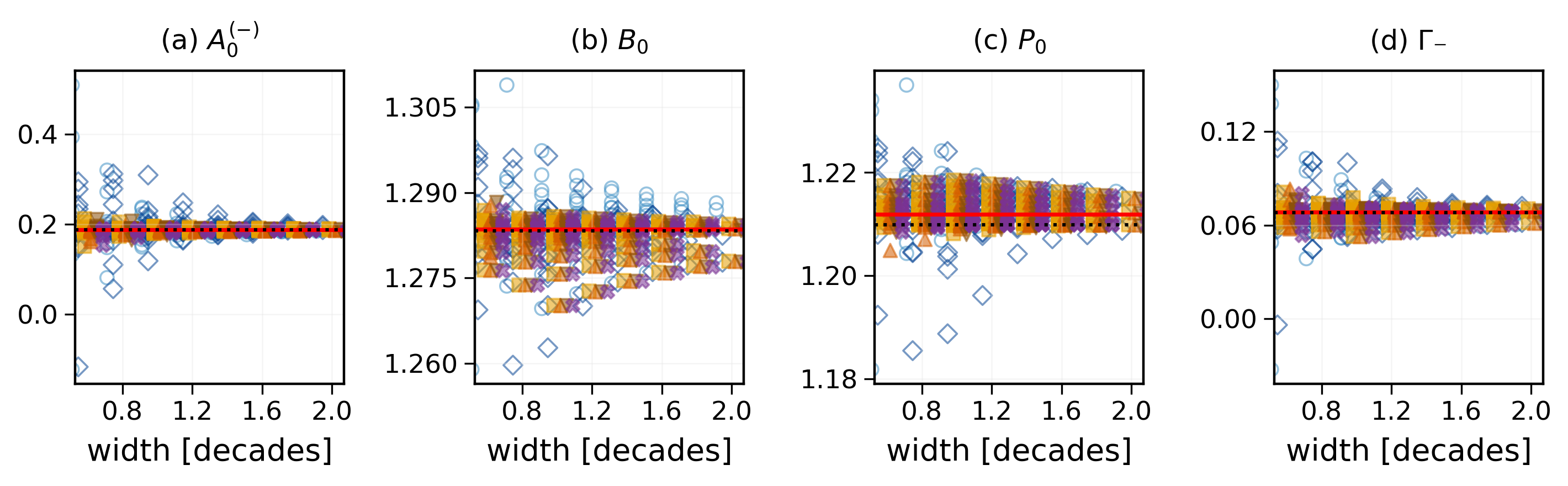}
  \caption{Dependence of fixed-exponent fit amplitudes on the width of the fitting window (in decades), for all $\valNfits$ combinations of window and truncation order. \BWDataKeyRef\ At each window width, the different datasets are slightly offset horizontally for improved legibility. Black dotted lines mark the exact reference values for $A_0^{(-)}$, $B_0$, and $P_0$; in the $\Gamma_-$ panel, the black dotted line corresponds to the Shchur--Janke reference value $0.0681(1)$~\cite{SHCHUR2010491}. Red solid lines indicate the median estimate at each window width. The spread of estimates decreases as the window width exceeds approximately one decade.}
  \label{fig:windowscan}
\end{figure*}

Figure~\ref{fig:windowscan} further elucidates the sensitivity of the fixed-exponent estimates to the choice of fitting window. It is observed that windows narrower than one decade produce the most extreme outliers; for example, estimates of $A_0^{(-)}$ range between $-0.12$ and $0.51$, and negative values for $\Gamma_-$ appear at the narrowest window width of $0.6$ decades. As the window width increases beyond approximately one decade, the estimates cluster more tightly around the exact reference values. The dataset-level uncertainties reported in Table~\ref{tab:amplitudes} are deliberately conservative, as they encompass the full spread of results from all $\valNfits$ window and truncation combinations, including those from the narrowest windows.

\subsection{Combined amplitude estimates}
\label{sec:results:combined}

To obtain robust, consolidated estimates of the critical amplitudes, the five independent datasets described in Sec.~\ref{sec:methods:datasets} were combined with equal statistical weighting. The uncertainty associated with each combined estimate was computed by adding in quadrature two contributions: the sample standard deviation calculated across the different datasets and the median dataset-level uncertainty (derived from the window-scan procedure). This approach ensures that both inter-dataset variability and intrinsic fitting uncertainties are properly accounted for.

The resulting combined amplitude estimates are summarized as follows:
\begin{align}
  B_0 &= \valBzero,      &  P_0 &= \valPzero,      \label{eq:B0P0_result}\\
  A_0^{(-)} &= \valAzeroMinus, & E_1^{(-)} &= \valEoneMinus, \\
  \Gamma_- &= \valGammaMinus,  & \Gamma_+ &= \valGammaPlus .
\end{align}
For the disordered phase, the combined estimates yield $A_0^{(+)}=\valAzeroPlus$ and $E_1^{(+)}=\valEonePlus$. These disordered-phase amplitudes are characterized by larger associated uncertainties, a feature that is further elaborated in Sec.~\ref{sec:results:collapse}.

A direct comparison with the theoretically exact values—$B_0=1.283265$, $P_0=1.209874$, $A_0=0.187788$, and $E_1=1.278376$—reveals excellent agreement within the reported uncertainties for all four energy and specific-heat amplitudes. Specifically, the combined estimates for $B_0$ and $P_0$ lie slightly above their exact counterparts: $B_0$ is higher by $0.06\%$, corresponding to $0.3$ times its uncertainty, while $P_0$ exceeds the exact value by $0.22\%$, or $0.9$ times its uncertainty. For the susceptibility amplitudes, the combined results for $\Gamma_-$ and $\Gamma_+$ are consistent with the reference values $\Gamma_-=0.0681(1)$ and $\Gamma_+=0.265(5)$ \cite{SHCHUR2010491}. Furthermore, an analysis of the Wolff algorithm's size dependence for $\Gamma_-$ indicates minimal variation with lattice size: the estimated values are $0.0688$, $0.0685$, $0.0686$, and $0.0679$ for $L=162$, $243$, $363$, and $486$, respectively, suggesting good convergence in this observable.

\subsection{Lattice-size dependence and correction-limited amplitude estimation}
\label{sec:results:collapse}

As the lattice size $L$ increases, the characteristic finite-size scale $\abs{t}_{\min}$ shifts to smaller values of the reduced temperature variable $x$, effectively extending the accessible range of the scaling regime. Within the majority of the reference fitting window, the effective exponents and effective amplitudes computed from the four Wolff datasets (at different $L$) exhibit remarkable agreement, as demonstrated in Figs.~\ref{fig:noplateau} and~\ref{fig:effamp}. This consistency implies that the dominant contributions to the observed deviations from asymptotic scaling behaviour arise from correction-to-scaling terms that are independent of the system size $L$.

The primary mechanism through which increasing $L$ influences the free-exponent fit results is via the selection of the fitting window. As larger lattices permit access to smaller $x$ values, the plateau-search algorithm may extend the fitting window down to $x\approx2\times10^{-4}$, which approaches the finite-size threshold $\abs{t}_{\min}$. For the magnetization amplitude $B_0$, this effect manifests as a systematic reduction in the free-fit deviation with increasing $L$: the deviation decreases from $6.4\%$ at $L=162$ to $5.4\%$, $4.7\%$, and $4.1\%$ at $L=243$, $363$, and $486$, respectively. Despite this improvement, these free-fit deviations remain substantially larger than those obtained from the fixed-exponent methodology. In contrast, the free-fit deviations for the polarization amplitude $P_0$ do not display a clear monotonic trend with $L$. For the MCPA data at $L=162$, the free-fit deviation for $B_0$ reaches $7.9\%$, whereas the combined fixed-exponent estimate aligns closely with the exact theoretical value.

The disordered-phase amplitudes $A_0^{(+)}$ and $\Gamma_+$ are associated with the largest window-scan uncertainties. This increased uncertainty stems from the relatively slow variation of their effective amplitudes across the fitting window, which results in weaker constraints on the correction-to-scaling parameters when using narrow windows. Moreover, the widest windows employed in the analysis extend to $x=0.2$, where higher-order correction terms begin to play a non-negligible role. It is noteworthy that Shchur and Janke adopted an even broader fitting range ($\abs{t}<0.77$) for their disordered-phase fits, extending well beyond the $x\le0.2$ limit used in the present study~\cite{SHCHUR2010491}.

\subsection{Determination of leading correction coefficients}
\label{sec:results:corrections}

The leading correction-to-scaling coefficients, denoted as $c_1$, were extracted by performing two-term fixed-exponent fits to the data within the reference window. Table~\ref{tab:corrections} presents the resulting coefficients for each observable.

\begin{table}[!ht]
\caption{Leading correction coefficients obtained from two-term fixed-exponent fits performed over the reference window $x\in[3\times10^{-3},5\times10^{-2}]$. The coefficient $B_1$ is known exactly, $P_1$ is determined in this work, and the reported $A_1$ values represent effective two-term coefficients that implicitly absorb the contributions of higher-order omitted terms. These effective coefficients are directly comparable to the values presented in Tables~I and~II of Ref.~\cite{SHCHUR2010491}.}
\label{tab:corrections}
\begin{ruledtabular}
\begin{tabular}{lrrrr}
\hline
Dataset & $B_1$ & $P_1$ & $A_1^{(-)}$ & $A_1^{(+)}$ \\
\hline
MCPA $L=162$ & -0.5462 & -0.2504 & -0.4190 & -0.2902 \\
MCPA $L=243$ & -0.5295 & -0.2328 & -0.4818 & -0.1853 \\
Wolff $L=162$ & -0.5272 & -0.2315 & -0.4414 & -0.2895 \\
Wolff $L=243$ & -0.5255 & -0.2299 & -0.4414 & -0.2635 \\
Wolff $L=363$ & -0.5253 & -0.2297 & -0.4401 & -0.3134 \\
Wolff $L=486$ & -0.5254 & -0.2299 & -0.4425 & -0.3129 \\
\hline
exact & -0.5898 & -- & -0.3884 & -0.3884 \\
\hline
\end{tabular}
\end{ruledtabular}
\end{table}

It is crucial to recognize that in a two-term fit, the extracted coefficient $c_1$ acts as an \textit{effective} parameter, encapsulating the cumulative effect of all higher-order terms that are not explicitly included in the model. The magnetization provides a particularly clear illustration of this effect. While the exact value of $B_1$ is known to be $-0.5898$, the combined two-term fit yields an effective coefficient of $B_1=\valBone$, which is smaller in magnitude. This finding is consistent with the results of~\cite{SHCHUR2010491}, who reported values of $-0.582(6)$ for $L=162$ and $243$, also exhibiting a reduced magnitude compared to the exact value. A similar pattern is observed for the specific-heat coefficients. The combined two-term estimates for the specific heat are $A_1^{(-)}=\valAoneMinus$ and $A_1^{(+)}=\valAonePlus$, which bracket the exact coefficient $A_1=-0.388$. The influence of the omitted second-order term $A_2\,t\abs{t}^{-2/3}$ (from Eq.~\eqref{eq:C_series}) on the effective amplitude $A_{\rm eff}$ is given by $A_2\operatorname{sgn}(t)\abs{t}$. Given that $A_2>0$, this term reduces the effective slope in the ordered phase and increases it in the disordered phase—a behaviour that is fully consistent with the observed trends.

For the polarization, the fitted correction coefficients demonstrate excellent consistency across all six datasets and both simulation algorithms, with mutual agreement within $10\%$. Notably, the four estimates derived from the Wolff datasets agree within $1\%$. The final combined estimate for the polarization correction coefficient is:
\begin{equation}
  P_1 = \valPone .
  \label{eq:P1}
\end{equation}
A comparison with the theoretical prediction based on the equal-relative-correction assumption (Sec.~\ref{sec:model:exact}) reveals a significant discrepancy. The predicted value $B_1P_0/B_0=-0.556$ has a magnitude that is $2.4$ times larger than $\abs{P_1}$. When comparing like-with-like—that is, using the fitted (effective) value of $B_1$ instead of the exact one—the predicted ratio becomes $B_1P_0/B_0=-0.500$, whose magnitude is still $2.1$ times $\abs{P_1}$. Interpreting both coefficients as effective two-term parameters, the relative correction for the polarization ($P_1/P_0\approx-0.19$) is therefore less than half the magnitude of the relative correction for the magnetization ($B_1/B_0\approx-0.41$; the exact value is $-0.46$).

\begin{figure}[!ht]
  \centering
  \includegraphics[width=\columnwidth]{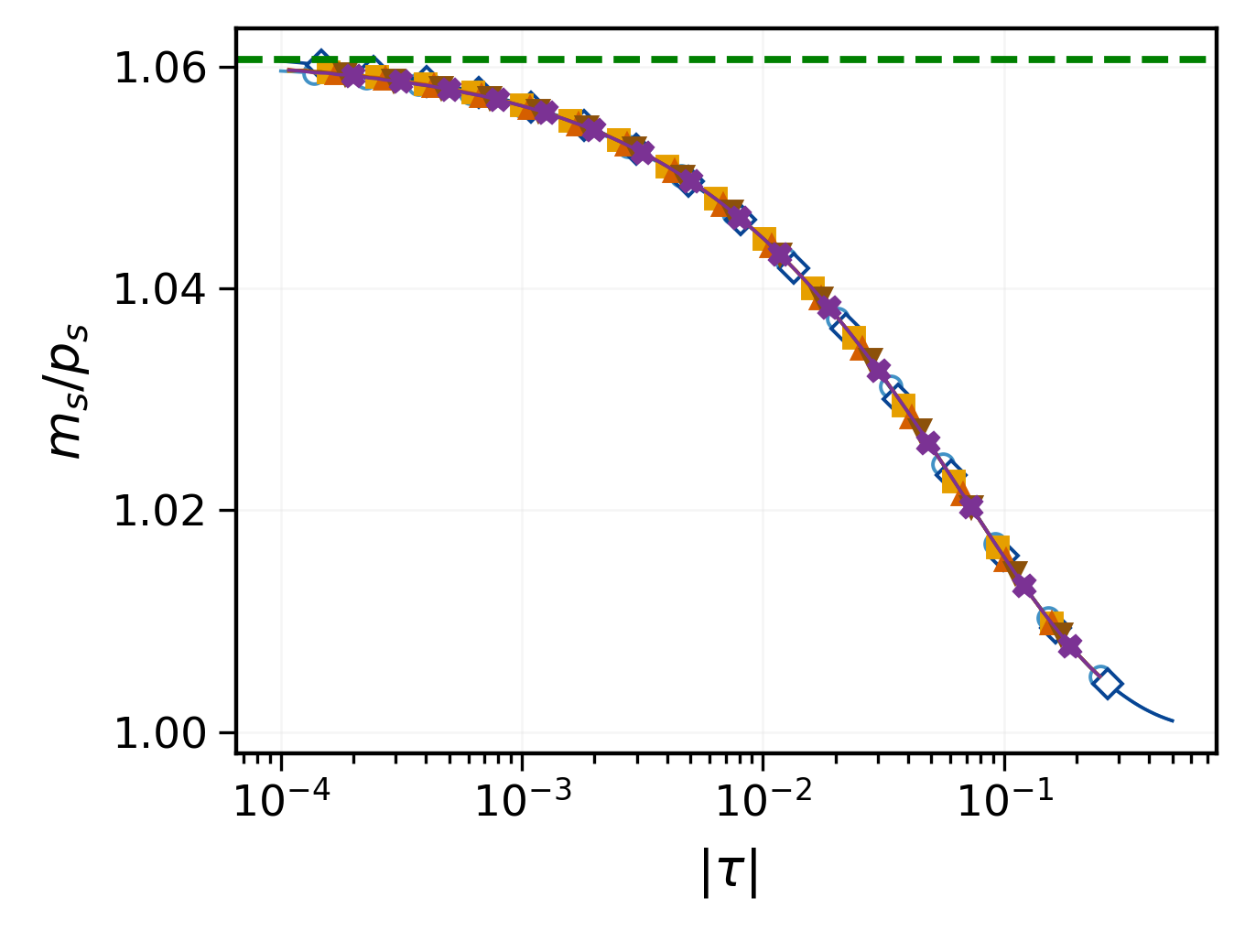}
  \caption{Measured ratio $m_s/p_s$ in the ordered phase as a function of $\abs{\tau}$. \BWDataKeyRef\ The green dashed line indicates the exact critical-limit value $B_0/P_0=3\sqrt{2}/4$. The non-constancy of the ratio provides direct evidence that the relative correction-to-scaling terms for $m_s$ and $p_s$ differ significantly.}
  \label{fig:ratio_ms_ps}
\end{figure}

Figure~\ref{fig:ratio_ms_ps} provides an independent, fit-free confirmation of this conclusion. If the relative corrections to $m_s$ and $p_s$ were identical, the ratio $m_s/p_s$ would remain constant across the entire scaling window. Instead, the measured ratio increases monotonically from approximately $1.00$ at large $\abs{\tau}$ towards the exact critical limit $3\sqrt{2}/4$ as $\abs{\tau}\to0$. This observed variation directly confirms that the relative correction terms for the two order parameters are distinct. Furthermore, the negative sign and magnitude of $P_1$ provide a quantitative explanation for the behaviour observed in Fig.~\ref{fig:p_vs_tau}: with $P_1/P_0\approx-0.19$, the sublattice polarization $p_s$ lies approximately $0.4\%$ below the leading-order term at $\abs{\tau}=3\times10^{-3}$ and about $2.6\%$ below it at $\abs{\tau}=5\times10^{-2}$.

\subsection{Universal amplitude ratios}
\label{sec:results:ratios}

Universal amplitude ratios provide a powerful means of probing critical behaviour, as they combine individual amplitudes in such a way that nonuniversal scale factors cancel out, leaving quantities that depend only on the universality class. In this study, we examine several key ratios: the ratio of the specific-heat amplitudes $A_0^{(+)}/A_0^{(-)}$, the susceptibility ratio $R_\chi=\Gamma_+/\Gamma_-$, and the specific-heat combinations $R_C^\pm=\alpha A_0^{(\pm)}\Gamma_\pm/B_0^2$, following the definitions established in Ref.~\cite{SHCHUR2010491}.

To assess the internal consistency of the fitted amplitudes, we also evaluate three additional checks: the ratio $E_1^{(+)}/E_1^{(-)}$, which is theoretically constrained to equal $A_0^{(+)}/A_0^{(-)}$ according to Eq.~\eqref{eq:E_from_A}; the ratio $B_0/P_0$, which reflects the relative normalizations of the sublattice magnetization and polarization; and the ratio $\Gamma_-/\Gamma_-^{(p)}$, which depends on the normalization of $p_s$ and $m_s$.

For the $q=4$ Potts universality class, Delfino and Cardy provided theoretical estimates using the two-kink form-factor approximation, yielding $R_\chi=4.013$, $R_C^-=0.00508$, and $R_C^+=0.0204$~\cite{DelfinoCardy1998}. These values serve as important benchmarks for our numerical results. Table~\ref{tab:ratios} presents the computed amplitude ratios derived from the fixed-exponent fit amplitudes for each individual dataset.

\begin{table*}[!ht]
\caption{Universal amplitude ratios computed from the fixed-exponent fit amplitudes. The definitions are $R_\chi=\Gamma_+/\Gamma_-$ and $R_C^{\pm}=\alpha A_0^{(\pm)}\Gamma_{\pm}/B_0^{2}$. In the reference row, the ratios $A_0^{(+)}/A_0^{(-)}$, $E_1^{(+)}/E_1^{(-)}$, and $B_0/P_0$ are exact theoretical values, while $R_\chi$ and $R_C^{\pm}$ correspond to the form-factor estimates from Ref.~\cite{DelfinoCardy1998}. The final row reproduces the data from Table~VI of Ref.~\cite{SHCHUR2010491} for comparative purposes.}
\label{tab:ratios}
\begin{ruledtabular}
\setlength{\tabcolsep}{4pt}\footnotesize
\begin{tabular}{lrrrrrrr}
\hline
Dataset & $A_0^{(+)}/A_0^{(-)}$ & $E_1^{(+)}/E_1^{(-)}$ & $B_0/P_0$ & $\Gamma_-/\Gamma_-^{(p)}$ & $R_\chi$ & $R_C^-$ & $R_C^+$ \\
\hline
MCPA $L=162$ & 1.02802 & 0.98615 & 1.05815 & 1.05370 & 4.20398 & 0.00484 & 0.02092 \\
MCPA $L=243$ & 0.94087 & 0.99935 & 1.05696 & 1.05867 & 3.79773 & 0.00523 & 0.01868 \\
Wolff $L=162$ & 1.00919 & 0.99208 & 1.05918 & 1.09325 & 3.95854 & 0.00525 & 0.02098 \\
Wolff $L=243$ & 1.00427 & 0.99940 & 1.05926 & 1.10347 & 3.91989 & 0.00521 & 0.02053 \\
Wolff $L=363$ & 0.98700 & 0.99735 & 1.05925 & 1.09913 & 3.92296 & 0.00521 & 0.02017 \\
Wolff $L=486$ & 1.02751 & 0.99845 & 1.05926 & 1.08699 & 3.96493 & 0.00516 & 0.02103 \\
\hline
reference & 1 & 1 & 1.06066 & -- & 4.013 & 0.00508 & 0.0204 \\
Shchur--Janke & 0.995(5) & -- & 1.061 & 1.11 & 3.9(1) & 0.00517(7) & 0.0201(5) \\
\hline
\end{tabular}

\end{ruledtabular}
\end{table*}

The combined estimate for the susceptibility ratio is $R_\chi=\valRchi$. This value shows excellent agreement with the form-factor prediction of $4.013$ and is also consistent with the estimate $3.9(1)$ reported in~\cite{SHCHUR2010491}. For the specific-heat combinations, we obtain $R_C^-=\valRCminus$ and $R_C^+=\valRCplus$. Both values are in good agreement with the theoretical form-factor estimates ($0.00508$ and $0.0204$, respectively) within the reported uncertainties.

The individual estimates of the ratio $A_0^{(+)}/A_0^{(-)}$ exhibit some scatter, varying between $0.94$ and $1.03$ around the exact theoretical value of $1$. The combined result, $\valRatioA$, is fully consistent with this exact value, demonstrating the robustness of our fitting procedure.

A particularly interesting case is the leading-amplitude ratio $B_0/P_0=\valRatioBP$, which can be compared directly with the exact theoretical prediction $3\sqrt{2}/4=1.06066$. The observed difference of $0.15\%$ corresponds to $3.4$ times the quoted uncertainty. Notably, this ratio is systematically lower than the exact value for every individual dataset, with all four Wolff datasets yielding values in the narrow range $1.0592$--$1.0593$. Although the individual amplitudes $B_0$ and $P_0$ each agree with their exact values within their respective uncertainties, the ratio $B_0/P_0$ exhibits significantly less scatter than either amplitude individually. This enhanced precision allows the ratio to resolve a subtle systematic offset between $P_0$ and $B_0$: while the combined $P_0$ lies $0.22\%$ above its exact value, $B_0$ is only $0.06\%$ above. One plausible explanation for this discrepancy is related to the choice of fit basis for the polarization: we adopt the correction powers derived from the magnetization series for the polarization fits. An exact polarization series or fits employing alternative correction bases would be required to rigorously test this hypothesis.

An alternative approach to determining $R_C^-$ involves constructing the ratio pointwise from observables measured at the same temperature, without relying on separate amplitude fits. In this method, we use $\tau$ as the common reduced temperature variable for both the thermal and magnetic scaling factors. While the relation $t=\tau+\mathcal{O}(\tau^2)$ implies that this choice modifies subleading analytic corrections, it preserves the critical limit. By combining the limiting forms $C\abs{\tau}^{\alpha}\to A_0$, $T\chi_-\abs{\tau}^{\gamma}\to\Gamma_-$, and $m_s^2\abs{\tau}^{-2\beta}\to B_0^2$, along with the scaling relation $\alpha+\gamma+2\beta=2$, we obtain the following expression:
\begin{equation}
  R_C^-(\abs{\tau}) = \alpha\,C\,(T\chi_-)\,\abs{\tau}^{2}/m_s^{2} ,
  \label{eq:RC_free}
\end{equation}
which converges to the desired universal ratio $\alpha A_0\Gamma_-/B_0^2=R_C^-$ as $\tau\to0$.

\begin{figure*}[!ht]
  \includegraphics[width=\textwidth]{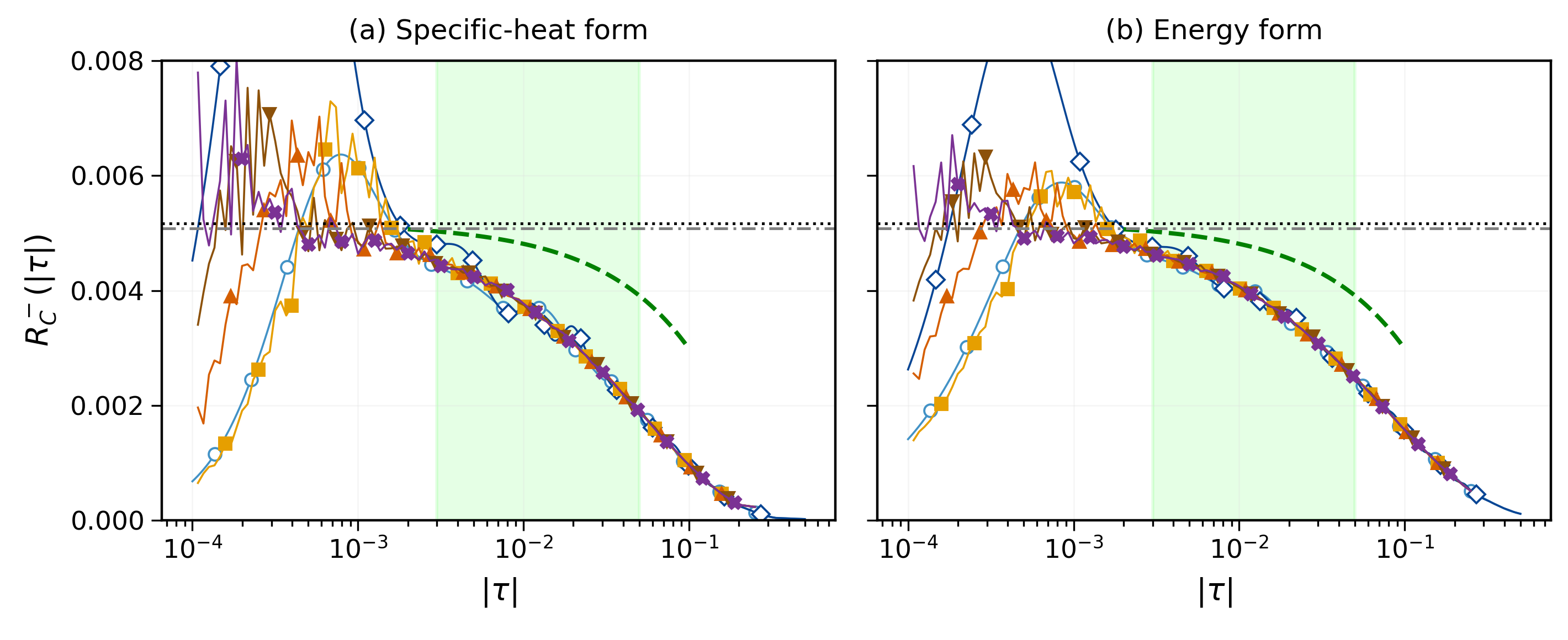}
  \caption{Pointwise representations of $R_C^-$ using the formulations in Eqs.~\eqref{eq:RC_free} and~\eqref{eq:RC_free_e}. Panel (a) shows the specific-heat form, while panel (b) displays the energy form. \BWDataKeyRef\ The green dashed curve provides a theoretical comparison, evaluating Eq.~\eqref{eq:RC_free} with the four-term Joyce series expansions for $C$ and $m_s$, assuming $t=\tau$ in the specific-heat factor, and using $T\chi_-=0.0681\abs{\tau}^{-7/6}$ from Ref.~\cite{SHCHUR2010491}. The black dotted line indicates the Shchur--Janke estimate $0.00517$, and the gray dash-dotted line marks the two-kink form-factor estimate $0.00508$~\cite{DelfinoCardy1998}. The green shaded band represents the reference fitting window $0.003\le\abs{\tau}\le0.05$, over which the two-term fits used to determine the limiting values are performed. Note that the MCPA $L=243$ curves extend beyond the plotted axes near $\abs{\tau}=10^{-3}$.}
  \label{fig:RCfree}
\end{figure*}

A second pointwise formulation replaces the specific heat with the energy. In the ordered phase, where $e<e_0=-\sqrt{2}$, the energy series implies $(1-\alpha)(e_0-e)\abs{t}^{-(1-\alpha)}/\Tc\to A_0$. Substituting $\abs{t}$ with $\abs{\tau}$ introduces a factor $(1-\tau)^{1-\alpha}$ that approaches unity near criticality. Combining this limit with the susceptibility and magnetization yields:
\begin{equation}
  R_C^-(\abs{\tau}) = \frac{\alpha(\alpha-1)}{\Tc}\,(e-e_0)\,(T\chi_-)\,
                      \frac{\abs{\tau}}{m_s^{2}} .
  \label{eq:RC_free_e}
\end{equation}
Since both $\alpha-1$ and $e-e_0$ are negative, the overall expression remains positive. The energy-based formulation has the advantage of using a first moment (the energy) rather than a variance (the specific heat), resulting in smaller point-to-point fluctuations (as illustrated in Fig.~\ref{fig:RCfree}). Shchur and Berche have argued that logarithmic corrections may cancel in this particular combination~\cite{ShchurBerche2008}, which could make the pointwise estimator valuable even for the $q=4$ Potts model where logarithmic terms are present.

To extract the limiting value of $R_C^-$, we extrapolate the pointwise forms to $\tau\to0$ using a two-term fit of the form $R_C^-(x)=c_0+c_1x^{2/3}$, where $x=\abs{\tau}$, over the reference window. We primarily report the result from the energy-based form due to its reduced statistical noise. The combined estimate from the five datasets yields $R_C^-=\valRCfree$. This value is $2.2\%$ below the form-factor estimate $0.00508$, representing a discrepancy of $2.5$ times its uncertainty. Given that the form-factor value itself is based on a two-kink approximation, this offset reflects the inherent limitations of both estimation methods. Importantly, the pointwise result is consistent with the value $\valRCminus$ obtained from the three independently fitted amplitudes.

\subsection{Susceptibility amplitudes and the interpretation of uncertainties}
\label{sec:results:limits}

\begin{figure*}[!ht]
  \includegraphics[width=\textwidth]{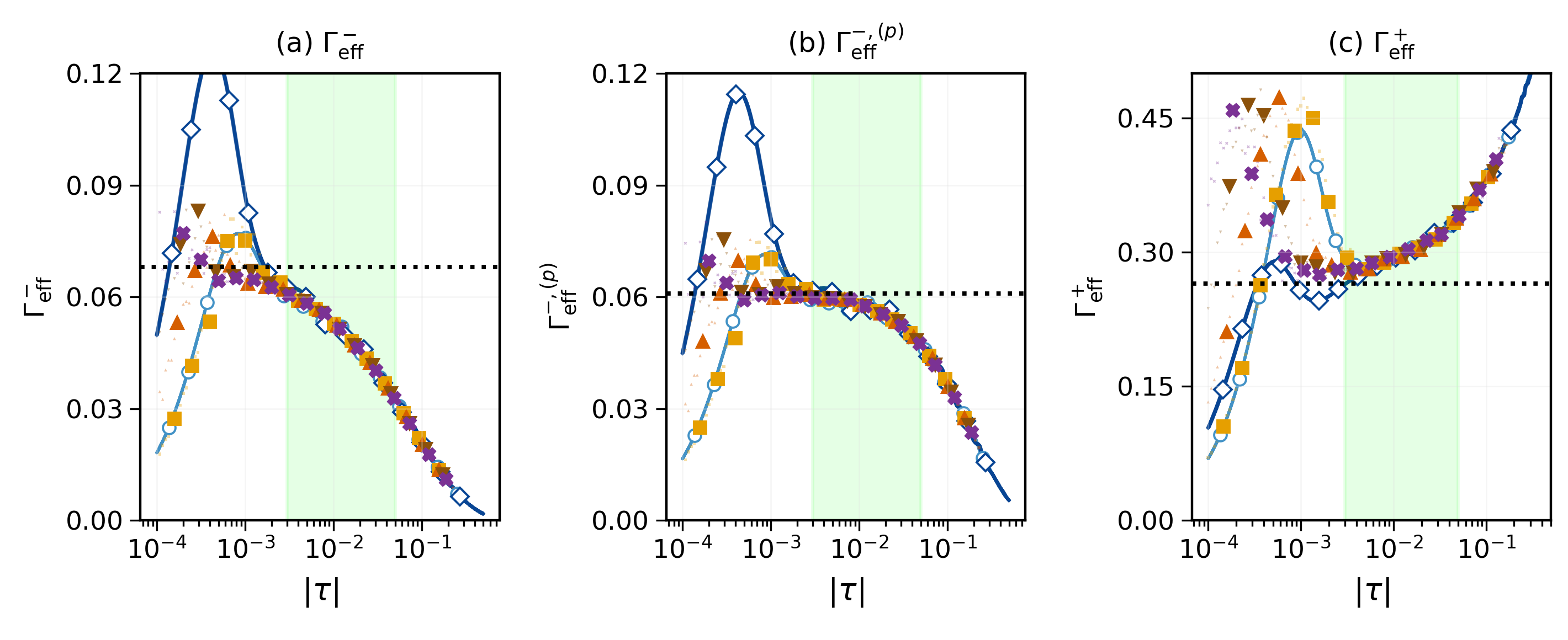}
  \caption{Effective susceptibility amplitudes defined as $\Gamma_{\rm eff}=T\chi\abs{\tau}^{7/6}$, constructed from $T\chi$ following the methodology of Ref.~\cite{SHCHUR2010491}. \BWDataKeyRef\ Panel (a) illustrates that the MCPA $L=243$ curve extends beyond the plotted range near $\abs{\tau}=5\times10^{-4}$. Black dotted lines indicate the reference values from Ref.~\cite{SHCHUR2010491}; it is important to note that their polarization susceptibility is based on $p$, while the curves shown here are based on $p_s$.}
  \label{fig:susceptibility}
\end{figure*}

Figure~\ref{fig:susceptibility} displays the effective susceptibility amplitudes, providing insight into the convergence properties of these quantities. Our estimate $\Gamma_-=\valGammaMinus$ is in excellent agreement with the value reported in~\cite{SHCHUR2010491}. However, our result for $\Gamma_-^{(p)}=\valGammaMinusP$ lies somewhat above their estimate of $0.061(1)$, leading to a ratio $\Gamma_-/\Gamma_-^{(p)}=\valRatioGamma$ that is lower than their reported value of $1.11$. This pattern is consistently observed for both simulation algorithms. As discussed in Sec.~\ref{sec:model:exact}, this discrepancy arises because in paper~\cite{SHCHUR2010491} define their polarization susceptibility using the global polarization $p$, whereas our analysis uses the sublattice polarization $p_s$. Consequently, these two values correspond to different physical observables; the comparison is provided here solely for orientation.

The uncertainties quoted for our amplitude estimates are designed to capture systematic sensitivity rather than pure statistical error. Specifically, they encompass the variability arising from different choices of the fitting window and truncation order, combined with the scatter observed among the five averaged datasets. It is crucial to recognize that these uncertainties do not include the variation expected from multiple independent simulation runs. Such variation would arise from multiple MCPA populations per system size and multiple Wolff Markov chains per temperature; in the current work, we have utilized only a single realization of each. The matched-size comparison presented in Sec.~\ref{sec:results:algorithms} provides an independent assessment of the agreement between the two algorithms.

Notably, the MCPA dataset at $L=243$ exhibits larger window-scan uncertainties compared to the $L=162$ dataset for the amplitudes $A_0^{(-)}$ and $\Gamma_-$. Moreover, the location of $t_{\rm peak}$ for the $L=243$ MCPA data breaks the trend observed for smaller system sizes, despite maintaining the same population size. More generally, when comparing equal system sizes, the MCPA window-scan uncertainties listed in Table~\ref{tab:amplitudes} exceed the corresponding Wolff uncertainties for 17 out of the 18 amplitude pairs; the sole exception is $P_0$ at $L=162$. To fully disentangle sampling errors from systematic deviations and to identify the origin of these differences, future work should employ multiple independent populations at each system size, with careful monitoring of the family-size distribution.

\section{Summary}
\label{sec:summary}

In this work, we leveraged the exactly known critical amplitudes of the two-dimensional Baxter--Wu model as a rigorous benchmark to evaluate and compare two distinct simulation algorithms—the MCPA (microcanonical population annealing) and the sublattice-freezing Wolff cluster algorithm—alongside two complementary fitting methodologies: the free-exponent and fixed-exponent approaches.

A direct comparison of the two algorithms at matched system sizes reveals strong overall agreement for key thermodynamic observables. Specifically, the relative root-mean-square differences between MCPA and Wolff results are $\valCrossE\%$ for the energy, $\valCrossMs\%$ for the sublattice magnetization $m_s$, and $\valCrossPs\%$ for the sublattice polarization $p_s$. Larger discrepancies are observed for fluctuation-sensitive quantities: the specific heat differs by $\valCrossC\%$, and the susceptibility by $\valCrossChi\%$. This hierarchy of differences is physically meaningful and aligns well with theoretical expectations: it is consistent with the first-order error decomposition outlined in Eq.~\eqref{eq:mcpa_observable_error} and with the well-known tendency for errors to be amplified when computing variances (such as those underlying specific heat and susceptibility). Following fixed-exponent fitting, both algorithms produce amplitude estimates that are in excellent agreement with the exact theoretical values. However, at comparable system sizes, the MCPA method generally yields larger window-scan uncertainties for most amplitudes compared to the Wolff algorithm, highlighting differences in systematic sensitivity between the two approaches.

The importance of fixing the critical exponent in amplitude fits—when its exact value is known—is clearly demonstrated by our results. The exact Joyce series for the Baxter-Wu model reveals that the local effective exponent exhibits a monotonic drift as a function of the scaling variable. This drift arises because the corrections to scaling form a series in powers of $t^{1/3}$, with coefficients of order unity, dominated by a term of relative order $t^{2/3}$. Consequently, any plateau-search procedure will tend to accept fitting windows where the effective exponent is drifting, leading the free-exponent fit to misinterpret this drift as an error in the exponent itself. This misattribution introduces a systematic bias in the fitted amplitudes. Quantitatively, the free-exponent approach yields a median deviation of $17\%$, with the deviation for each individual amplitude closely tracking the magnitude of its relative correction term. In stark contrast, the fixed-exponent fit—which incorporates the analytically known correction series—reduces the median deviation dramatically to just $0.37\%$. Furthermore, while increasing the system size $L$ improves convergence for some quantities, the free-fit bias for the magnetization amplitude $B_0$ decreases only slowly with $L$, and no clear trend is observed for the polarization amplitude $P_0$, underscoring the persistent systematic nature of the bias introduced by the free-exponent method.

Using the robust fixed-exponent fitting framework, we obtained highly accurate combined amplitude estimates. The results $B_0=\valBzero$, $P_0=\valPzero$, $E_1^{(-)}=\valEoneMinus$, and $A_0^{(-)}=\valAzeroMinus$ are all in excellent agreement with their exact theoretical counterparts. Similarly, the susceptibility amplitudes $\Gamma_-=\valGammaMinus$ and $\Gamma_+=\valGammaPlus$ align well with the reference values reported by Shchur and Janke~\cite{SHCHUR2010491}. The universal amplitude ratios—$R_\chi=\valRchi$, $R_C^-=\valRCminus$, and $R_C^+=\valRCplus$—are consistent with the theoretical form-factor estimates derived by Delfino and Cardy~\cite{DelfinoCardy1998}. Notably, the ratio $B_0/P_0=\valRatioBP$ lies $3.4$ standard uncertainties below its exact value of $3\sqrt{2}/4$. While this small systematic offset is statistically significant, it is not large in absolute terms and likely points to a subtle issue in the fitting procedure for $P_0$, potentially related to the choice of correction basis (i.e., using magnetization-based correction powers for the polarization fit). The leading polarization correction coefficient was determined to be $P_1=\valPone$, and its relative magnitude ($P_1/P_0\approx-0.19$) is found to be less than half that of the corresponding magnetization correction ($B_1/B_0\approx-0.41$), providing quantitative evidence for the differing strengths of scaling corrections in these two order parameters.

Several promising directions for future research naturally emerge from this study. First, implementing multiple independent MCPA populations at each system size—with careful monitoring of the family-size distribution—would allow the current window-scan uncertainties to be converted into genuine statistical errors. Such an approach would also help clarify the anomalous behaviour observed in the $L=243$ MCPA dataset, where the peak position $t_{\rm peak}$ deviates from the trend seen at smaller sizes. Second, the methodological framework developed here—particularly the fixed-exponent fitting strategy and the pointwise estimator for $R_C^-$—can be extended to the $q=4$ Potts model. In this more complex case, however, the fitting series would need to be augmented to explicitly include logarithmic correction terms, which are known to play a significant role in the critical behaviour of the $q=4$ Potts universality class.

\section*{Data Availability Statement}
The data and analysis code supporting this article are available from the
authors upon request.

\bibliography{references}

\end{document}